%% file: main.tex
\documentclass[twocolumn, superscriptaddress, amsmath, amssymb, pre, floatfix]{revtex4-1}
\usepackage[utf8]{inputenc}
\usepackage[T1]{fontenc}
\usepackage[tight]{subfigure}
\subfiguretopcaptrue
\input{mydiagrams}

\input{preamble}

\newcommand{\mS}{\mathcal{S}}
\newcommand{\Fref}[1]{Fig.~\ref{fig:#1}}
\newcommand{\OC}{\mathcal{O}}
\newcommand{\Expected}[1]{\mathbb{E}[#1]}
\newcommand{\phitilde}{\widetilde{\phi}}

\renewcommand{\Sref}[1]{Sec.~\ref{Sec:#1}}

\newcommand{\Cov}{\operatorname{Cov}}
\usepackage{comment}

\begin{document}

\title{Time-dependent branching processes: a model of oscillating neuronal avalanches}
\author{Johannes Pausch}
\email[]{jp634@cam.ac.uk}
\affiliation{Department of Mathematics and Centre for Complexity Science, Imperial College London, London SW7 2AZ, United Kingdom}
\affiliation{Department of Applied Mathematics and Theoretical Physics and St.~Catharine's College, University of Cambridge, Cambridge CB3 0WA, United Kingdom}
\author{Rosalba Garcia-Millan}
\email[]{garciamillan16@imperial.ac.uk}
\affiliation{Department of Mathematics and Centre for Complexity Science, Imperial College London, London SW7 2AZ, United Kingdom}
\author{Gunnar Pruessner}
\email[]{g.pruessner@imperial.ac.uk}
\affiliation{Department of Mathematics and Centre for Complexity Science, Imperial College London, London SW7 2AZ, United Kingdom}
\date{\today}

\begin{abstract}
Recently, neuronal avalanches have been observed to display oscillations,
a phenomenon regarded as the co-existence of a scale-free behaviour (the
avalanches close to criticality) and scale-dependent dynamics (the 
oscillations). Ordinary continuous-time branching processes with constant
extinction and branching rates are commonly used as models of neuronal
activity, yet they lack any such time-dependence.
In the present work, we extend a basic branching process by allowing the
extinction rate to oscillate in time as a new model to describe cortical
dynamics.  By means of a perturbative field theory, we derive relevant
observables in closed form. We support our findings by quantative comparison to
numerics and qualitative comparison to available experimental results.
\end{abstract}

\maketitle

\section{Introduction}
In the brain, electrical signals propagate between neurons of the cortical network through action potentials, which are spikes of polarisation in the membrane of the neuron's axon. These spikes have an amplitude of about $100$mV, typically last about 1ms \cite{Dayan2001} and can be recorded using micro-electrodes \cite{Plenz1996,Plenz1998}. In order to study the signaling in larger regions of neurons, multielectrode arrays, comprising about $60$ electrodes spread across $\approx4$mm$^2$, are used to capture the collective occurrence of spikes as local field potentials (LFPs). In this setting, the electrodes are extracellular and each is sensitive to electrical signals from several surrounding neurons \cite{Karpiak2002,BeggsPlenz:2003,BeggsPlenz:2004,Priesemann:2009,Priesemann2013}. 

The data of the LFP recordings are processed by putting them into time bins of a few milliseconds length and by introducing a voltage threshold. In addition, a refractory period is imposed to avoid counting large voltage excursions more than once. 
The details of processing can differ slightly between experiments \cite{BeggsPlenz:2003,BeggsPlenz:2004,Wagenaar2006,Priesemann:2009,Priesemann2013}. However, after processing, the data is a time series of two values for each electrode: on (detected signal above threshold)  and off (no detected signal or signal below threshold). A \textit{neuronal avalanche} is then defined as
a set of uninterrupted signals detected across the micro-electrode array. Each avalanche is both preceded and succeeded by at least one time bin where none of the electrodes detected a signal, defining the avalanche duration as the number of time bins where avalanche unfolds. Which and how many electrodes detect a signal varies during the avalanche  \cite{BeggsPlenz:2003,BeggsPlenz:2004}. The duration of avalanches typically ranges between a few milliseconds and 30ms \cite{Beggs:2008}. A prominent observable is the avalanche size, which is the total number of recorded signals during the avalanche. If there is only one electrode detecting a signal in each time bin of an avalanche, its size and duration are equal. However, the size of an avalanche is usually larger than its duration due to the simultaneous detection of signals by different electrodes. Histograms of the avalanche size show an apparent \textit{power-law distribution of sizes}, with common small avalanches and rare large avalanches \cite{BeggsPlenz:2003,BeggsPlenz:2004}, the fingerprint of scale-free phenomena. The exponent of the power law was observed to be $-3/2$ in \cite{BeggsPlenz:2003}. However, its exact value was found to be significantly sensitive to the choice of time bin size \cite{Priesemann2013} and spatial sub-sampling of the neural network \cite{Priesemann:2009}.

The observation of this power-law distribution led to the hypothesis that neuronal avalanches can be modelled adequately by models in the class of self-organized-criticality  (SOC) \cite{BeggsPlenz:2003,Priesemann2013,Brochini2016} or models of critical branching processes \cite{BeggsPlenz:2003,Haldeman2005,Williams-Garcia2014,Wilting2019}, because both are avalanche models showing power-law distribution of the avalanche size \cite{Pruessner:2012:Book,Garcia-Millan2018}. In this article, we focus on modelling neuronal avalanches as branching processes. 

A \textit{branching process} is a stochastic process in which a particle can either randomly create identical copies of itself or spontaneously go extinct,
triggering an avalanche of particles \cite{Watson:1875,Harris:1963,AthreyaNey:1972,Pazsit:2007,Williams:2013,Marzocchi:2008,Lee:2004,Simkin2010,Williams:2013,durrett2015branching,GleesonDurrett:2017,seshadri2018altered,Garcia-Millan2018}. The particle number $N$ in the branching process is interpreted as the number of electrodes that detect a signal. The creation of a new particle is interpreted as
the change of an electrode from no detection to the detection of a signal
 \cite{BeggsPlenz:2003}, while the extinction of a particle corresponds to the change from detection to no detection \cite{Poil2008,Priesemann2013,Wilting2019}. Branching processes display a phase transition from asymptotic extinction with probability $1$ to asymptotic survival  with a positive probability, depending on the average number of created particles. At the critical point, the avalanche size of a branching process is power-law distributed
 \cite{Garcia-Millan2018}.

Recently, there has been evidence that neuronal activity, when compared to branching processes, is not at criticality but in a reverberating regime close to criticality \cite{Wilting2018,Wilting2018b,Wilting2019}. The reason for the strong interest in the system's distance to a critical point is the criticality hypothesis, which states that information processing in the brain might be optimized by the cortical network being close to a critical point \cite{Haldeman2005,Beggs:2008,Williams-Garcia2014,Timme2016,Wilting2019b}. However, fitting power laws is notoriously difficult \cite{Goldstein2004,Wilting2018b} and other means of verifying the criticality of the neuronal avalanches are essential. For this reason, the \textit{avalanche shape}, defined as the average temporal profile of the avalanches, has received more attention in recent years. It is debated whether, at criticality, the neuronal avalanche shape takes the universal form of an inverse parabola  \cite{Papanikolaou2011,Beggs2012,Laurson2013,Rybarsch2014,GleesonDurrett:2017,Garcia-Millan2018,MillerYuPlenz:2019}, which is the case of a critical branching process. The universality of this shape has been particularly challenged in \cite{MillerYuPlenz:2019} by observations of oscillations that modulate the avalanche shape. In \cite{MillerYuPlenz:2019}, the oscillations are identified as $\gamma$-oscillations, which are a particular frequency band of brain waves between $30$ and $100$Hz. Brain waves are electric oscillations spanning the entire brain that can be recorded using  electroencephalography (EEG) \cite{Berger1929,Buzsaki2004}. The oscillations are organized into bands covering frequencies between 0.05Hz (slow 4 band) and 500Hz (ultra fast band). Their power spectrum follows approximately a $1/f$ distribution \cite{Penttonen2003}.

The observation of oscillations in the EEG activity on the one hand, and the observation of scale-free avalanches of LFP activity on the other hand raises the question of how these two seemingly incompatible descriptions of the same phenomena can be reconciled. The experimental data in \cite{MillerYuPlenz:2019} indicates that $\gamma$-oscillations modulate the avalanche shape. Can branching processes incorporate oscillations as well? Will those oscillations modulate the avalanche shape, widely regarded as a universal feature?

To answer these questions, we extend in this paper the field theory in \cite{Garcia-Millan2018} by
incorporating oscillatory extinction rates, Sec.~\ref{Sec:Birth-Death-Model}. We then calculate observables such as moments of the particle number, Sec.~\ref{Sec:Moments}, its covariance, Sec.~\ref{Sec:Covariance}, survival probability, Sec.~\ref{Sec:SurvProb}, and the avalanche shape (or temporal profile), Sec.~\ref{Sec:Shape} and compare them qualitatively to experimental results from \cite{MillerYuPlenz:2019}. 
We conclude in Sec.~\ref{Sec:Conclusion}.

\section{Model}
\label{Sec:Birth-Death-Model}

\begin{figure}
    \centering
    \includegraphics[width=\columnwidth]{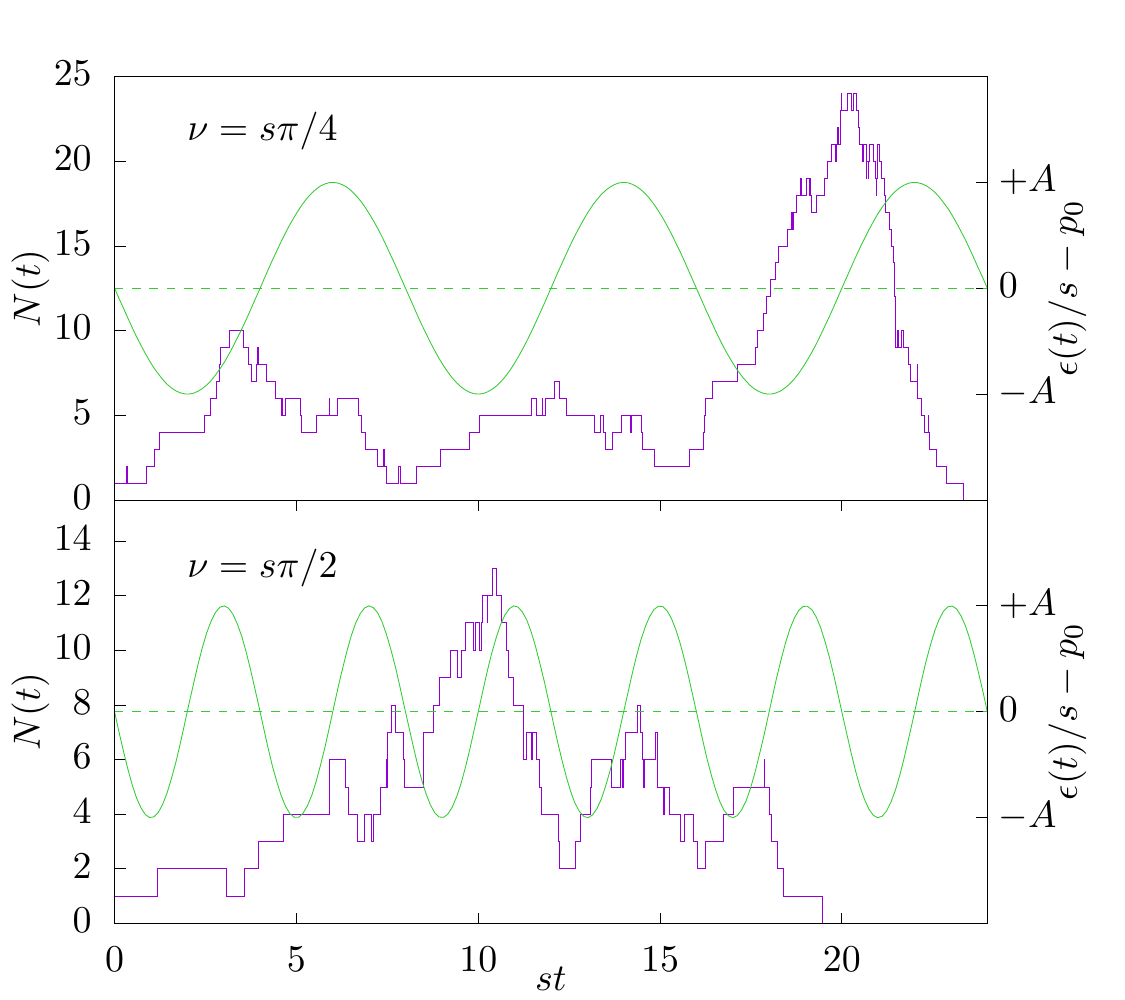}
    \caption{Two example trajectories of the number of particles $N(t)$ (purple, left ordinate) of branching processes with
    $r=0$ and periodically varying extinction rate $\epsilon(t)=s(p_0-A\sin(\nu t))$, \Eref{extinction-rate}. Both trajectories were initialized with one particle at $t_0=0$. The perturbation of the extinction rate, $-A\sin(\nu t)$ with $\nu/s\in\{\pi/4,\pi/2\}$, $A=0.5$ is shown in green
    (right ordinate). When the extinction rate is lower, branching is effectively promoted. 
    }
    \label{fig:trajectory}
\end{figure}

A branching process in continuous time $t$ can be regarded as a reaction of a single particle   type $B$ that splits into $K\in\mathbb{N}_0$ copies of itself,
\begin{equation}\elabel{Reaction}
B\xrightarrow{s}KB \, ,
\end{equation}
with Poissonian rate $s$ \cite{Harris:1963,AthreyaNey:1972,Pazsit:2007}. The particle number at time $t$, which is also called population size, is denoted by $N(t)$. In the event of a reaction, the population size increases 
by $K-1$ particles.
The waiting time for an individual particle
to undergo any such reaction is
 exponentially distributed with rate $s$. 
The number of offspring $K$ 
may be a  random variable itself, with probability
distribution $P(K=k)=p_k$. 
We call the reaction in \eref{Reaction} a branching event 
if
$K>1$ and an extinction event if $K=0$. 
Throughout the present work we 
initialize the process 
with one particle at time
$t_0=0$, so that $N(0)=1$.
The totality of a realisation of a branching process, from the initialization until the termination
after which the population size remains $0$ indefinitely is referred to as an avalanche.

In \cite{Garcia-Millan2018}, we 
introduced a 
Doi-Peliti field theory for the continuous-time branching process with time-independent but arbitrary offspring distribution $p_k$. We refer to such a branching process as \textit{standard branching process}. The action functional of this field theory is
\begin{equation}\label{eq:birth-death-action}
    \mathcal{A}_0[\phi,\widetilde\phi]=
    \int\left\{\widetilde\phi\left(-\frac{\plaind}{\plaind t}-r\right)\phi+\sum\limits_{j=2}^\infty q_j\widetilde\phi^j\phi
    \right\}\plaind t \, ,
\end{equation}
with 
\begin{equation}\elabel{def_q_j_and_r}
r=s\left(1-\mathbb{E}[K]\right)
\text{ and } 
q_j=s\sum\limits_{k=j}^\infty 
\binom{k}{j}
p_k\ ,
\end{equation}
and time-dependent Doi-shifted annihilation and creation fields 
$\phi(t)$ and $\widetilde\phi(t)$ 
respectively.
The parameter $r$ is called the mass in 
the context of field theories and is, according to \Eref{def_q_j_and_r}, 
closely linked to the 
first moment of the 
offspring distribution
$\mathbb{E}[K]$. Traditionally \cite{Harris:1963}, the branching process is called \emph{sub-critical} if 
$\mathbb{E}[K]<1$ (and thus $r>0$),
it is called \emph{critical}
if $\mathbb{E}[K]=1$ (and thus $r=0$),
and \emph{super-critical} if
$\mathbb{E}[K]>1$ (and thus $r<0$).

The time-scale of this branching process
is set by $s$, the rate with which
\emph{any} single-particle event takes place.
The rate with which any particle 
spontaneously disappears, the 
 extinction rate, is thus
$\epsilon = sp_0$.

In the following we will focus on 
binary branching processes, \ie 
$p_k=0$ for all $k$ except $k=0$ and $k=2$,
so that $p_2$ and $p_0=1-p_2$ in \Eref{def_q_j_and_r} are determined by $r/s=1-2p_2$. The rest of parameters follow then immediately,
such as $\mathbb{E}[K]=2p_2$, 
$q_2/s=p_2=1/2(1-r/s)$ and $q_j=0$ for $j\ge3$,
as well as bounds such as $s\ge r+q_2\ge0$. 
In particular $r=\epsilon - q_2$, so that the branching
process is critical if $q_2$ balances
the extinction rate. Extensions to
branching processes with other time-independent offspring distributions are
straight forward \cite{Garcia-Millan2018}.

Our extension to this model \eref{birth-death-action} consists 
of a \textit{time-dependent}, oscillating
extinction rate 
\begin{equation}
    \epsilon(t)=s\big(p_0-A\sin(\nu t)\big) \, ,
    \label{eq:extinction-rate}
\end{equation}
where 
$\nu$ is
the frequency of the oscillations. The dimensionless amplitude $A$ is the small
parameter of the field theoretic 
perturbation theory 
about $A=0$. In the following, $r\ge0$ is
considered.
The magnitude $|A|$ is bounded by $p_0$ because $\epsilon(t)$, as an extinction rate, has to be non-negative at all times. If the amplitude $A$ is positive, 
then the extinction rate is initially suppressed, favouring more
branching events. In the field theory, this extension amounts to 
adding the term  (details in App.~\ref{Sec:DerivationAction}),
\begin{align}\elabel{A-term}
    As\int\left\{\widetilde\phi(t)\phi(t) \sin(\nu t)\right\}\plaind t \, ,
\end{align}
to the original action $\mathcal{A}_0$,
\Eref{birth-death-action},
where $r=sp_0-sp_2$ is still the “bare mass”. 
Our extension can be summarised as
a standard branching process where binary branching takes place with rate $q_2=sp_2$ and extinction with rate $\epsilon(t)$, \Eref{extinction-rate}. We retain the parameterisation $p_0+p_2=1$. 

It will turn out that the perturbation
remains noticeable at all times in the process in
all observables considered. In 
particular, even at criticality, the shape of the avalanche, Sec.~\ref{Sec:Shape}, carries a clear signature of the
oscillations despite its expected 
universality \cite{Papanikolaou2011}. 
The model
and the analysis in the present work 
therefore provide an explanation for
the shape of the temporal profile
of neuronal avalanches recently reported by
Miller \etal \cite{MillerYuPlenz:2019}.

\Fref{trajectory} 
shows two example trajectories 
of the population size $N(t)$,
together with the perturbation of the extinction rate, 
$\epsilon(t)/s-p_0=-A\sin(\nu t)$.
In all figures, the latter is 
shown in green with the 
ordinate on the right.
All data in this work is 
presented in
dimensionless form, in particular time
as $st$.

In the following, we state the central results, whose field-theoretic
derivation is relegated to the Appendix. In particular, we consider the avalanche shape at criticality, Sec.~\ref{Sec:Shape}, which is a common observable in LFP recordings of the brain \cite{BeggsPlenz:2003,BeggsPlenz:2004,Priesemann:2009,Priesemann2013} and the covariance, Sec.~\ref{Sec:Covariance}, which recently gained more attention in neuroscience as a tool to estimate the system's distance to the critical point \cite{Wilting2018,Wilting2018b,Wilting2019}.

\section{Moments}
\label{Sec:Moments}
\subsection{First Moment}
\label{Sec:first_moment}
Since the extinction rate varies periodically, we expect that the first moment varies accordingly. As shown in appendix~\ref{Sec:FirstMomentDerivation}, the expected
particle number is,
\begin{subequations}
\begin{align}
    \mathbb{E}[N(t)]&=\Theta(t)\text{exp}
    \left(-rt+\int\limits_{0}^t As\sin(\nu t')\plaind t'\right) \\
    &
    =\Theta(t)\text{exp}
    \left(-rt-\frac{As}{\nu}(\cos(\nu t) - 1) \right) \, , \label{eq:exact-first-moment}
\end{align}
\end{subequations}
where $\Theta(t)$ is the Heaviside function reflecting the initialization at $t_0=0$, henceforth dropped from all
expressions.
This result is consistent with the known result of an inhomogenous Poisson process with a time-dependent event rate \cite{Kingman1992}.

\begin{figure}
    \centering
    \includegraphics[width=\columnwidth]{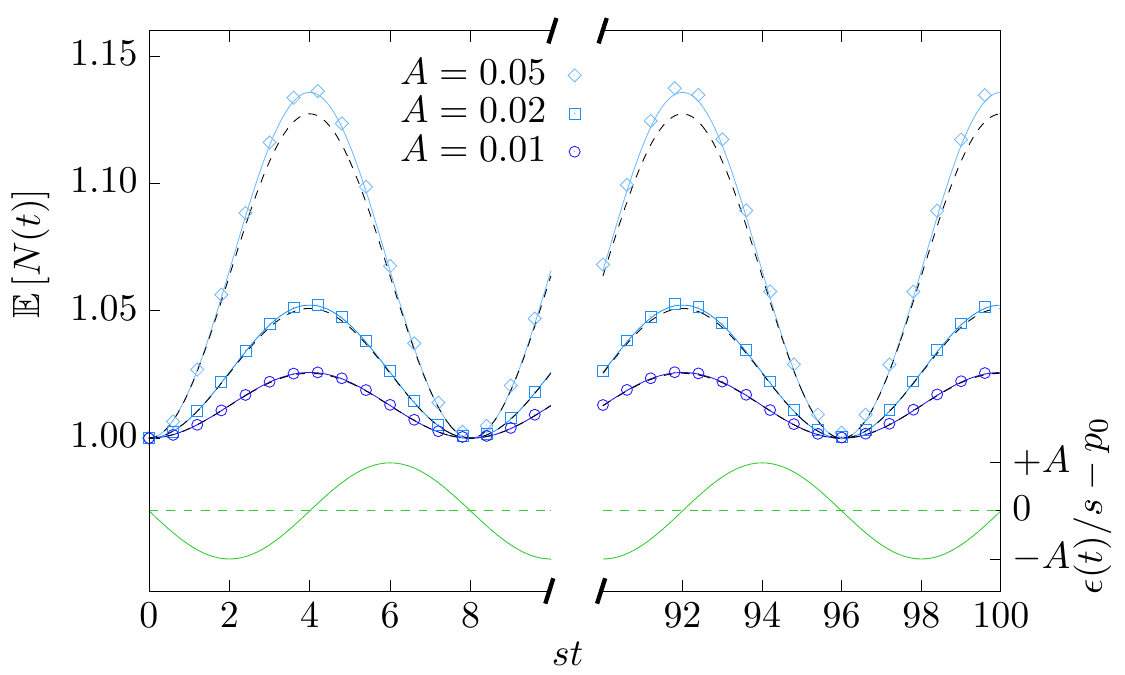}
    \caption{Expected particle number $\mathbb{E}[N(t)]$ for $A\in\{0.01,0.02,0.05\}$, $\nu/s=\pi/4$ and $r=0$. The system was initialized with a single particle at $t_0=0$. 
    Symbols: simulation results using $10^9$ realizations. Full blue lines: exact analytic predictions, Eq.~\eqref{eq:exact-first-moment}. Dashed black lines: analytic approximation 
    $\Expected{N(t)}=g_{1}^{(0)}(0,t)+Ag_{1}^{(1)}(0,t) + \OC(A^2)$, see
    Eqs.~\eqref{eq:moment-approximation-notation},~\eqref{eq:moments-first-order-correction} and~\eref{gn_approx_of_N1}.
    The perturbation of the extinction rate, \Eref{extinction-rate}, is shown in green (right ordinate).}
    \label{fig:firstMoment}
\end{figure}

Fig.~\ref{fig:firstMoment} shows an estimate of the first moment 
based on Monte-Carlo simulations together with the analytical result \Eref{exact-first-moment} for
$r=0$.
In addition to this exact result, the dashed line in 
\Fref{firstMoment} shows the
first order approximation
$\Expected{N(t)}=g_1^{(0)}(0,t)+Ag_{1}^{(1)}(0,t)+\OC(A^2)$,
Eq.~\eqref{eq:moment-approximation-notation}, which is discussed below in Sec.~\ref{Sec:NMomentDerivation}.

Because the extinction rate is reduced during the first half-period, Eq.~\eqref{eq:extinction-rate}, branching dominates the process and the population size increases at first. 
During the second half-period, extinction dominates 
and the population size decreases, perfectly balancing, on
average, the creation of particles in the first half of the period.
According to \Eref{exact-first-moment} and \Fref{firstMoment}, the expected population size
$\Expected{N(t)}$ at $r=0$
is always equal or larger than unity, 
which is the expected number of particles of a standard branching process
at criticality. Thus, while the extinction rate is not shifted on average, the expected particle number is on average larger than in the process without oscillations.

Just like in a standard 
branching process, $\Expected{N(t)}$
converges to $0$ for $r>0$ and
diverges for $r<0$, indicating that the present process is 
critical at $r=0$. However,
unlike a standard 
branching process, the expected
population size never converges for $r=0$ as the
oscillations never cease.
It will turn out that the effective
mass acquires no shift
due to the perturbation.

\subsection{Second Moment}
\label{Sec:second-moment}
The second moment can be calculated in closed form from a convolution
integral involving the first moment only, App.~\ref{Sec:SecondMomentDerivation}. 
From \Esref{first_moment_general}, \eref{2nd_moment} and \eref{secondmom_from_firstmom},
\begin{align}\label{eq:exact-second-moment}
     &\mathbb{E}[N^2(t)]=
   \text{exp}\hspace{-0.1cm}\left(-rt-\frac{As}{\nu}\big(\cos(\nu t)-1\big)\right)\\
    &\hspace{-10pt}\times\left[1+2q_2\int\limits_{0}^{t}\text{exp}\hspace{-0.1cm}\left(-r(t-t')
    -\frac{As}{\nu}\big(\cos(\nu t)-\cos(\nu t')\big)\right)\plaind t'\right] \, .
    \nonumber
\end{align}
At $r=0$
the second moment is
to first order in $A$
\begin{multline}
    \mathbb{E}[N^2(t)] = 
    1+2q_2t+ \frac{As}{\nu} \Big[1+2q_2t +\frac{2q_2}{\nu}\sin(\nu t) \\
     - (1+4q_2t)\cos(\nu t)  \Big] +\mO\left( A^2\right)
    \elabel{first-order-second-moment}
\end{multline}
which diverges asymptotically linear in $t$. Because  $\Expected{N(t)}$ is bounded, the variance also diverges linearly in~$t$. 
\Fref{secondMoment} shows the ratio 
$\mathbb{E}[N^2(t)]/(1+2q_2t)$, which illustrates the deviation of 
the second moment at $A>0$ from that at $A=0$.
For large $t$, 
the second moment shows
a linear increase with an amplitude that oscillates mildly with period $\nu$,
so that 
$\mathbb{E}[N^2(t)]$
oscillates around the $A=0$ behaviour. For large $q_2t$ the ratio 
$\mathbb{E}[N^2(t)]/(1+2q_2t)$ is to leading order $1-(As/\nu)(2\cos(\nu t)-1)$.

\begin{figure}
    \centering
    \includegraphics[width=\columnwidth]{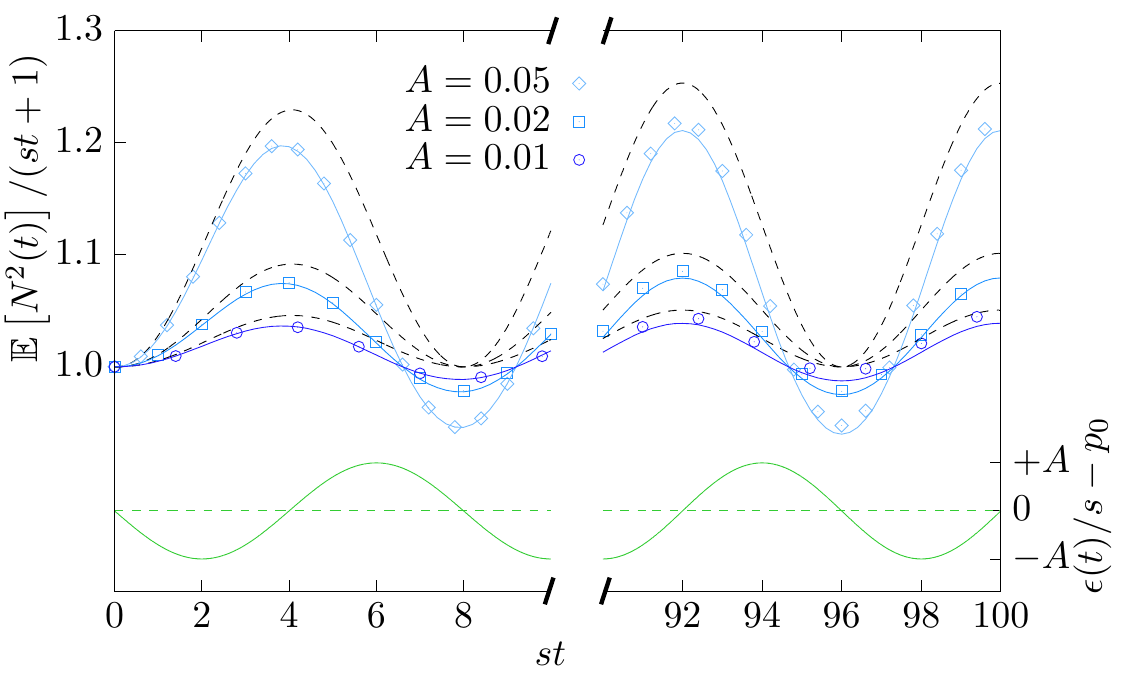}
    \caption{Rescaled second moment of the particle number for $A\in\{0.01,0.02,0.05\}$, $\nu/s=\pi/4$ and $r=0$. The second moment is rescaled by the second moment with constant extinction rate, $\mathbb{E}[N^2(t);A=0,r=0]=1+st$. The system was initialized with a single particle at $t_0=0$.
    Symbols: Simulations results using $10^7$ realizations. Straight blue lines: Exact analytic result, Eq.~\eqref{eq:exact-second-moment}. Dashed black lines: Approximation to first order in $A$, see  Eq.~\eqref{eq:moments-first-order-correction} and  \Eref{gn_approx_of_N2}.
    The perturbation of the extinction rate, \Eref{extinction-rate}, is shown in green (right ordinate).}
    \label{fig:secondMoment}
\end{figure}

Although the extinction is not shifted on average, the second moment is shifted on average due to the oscillations, see Fig.~\ref{fig:secondMoment}.

\subsection{n-th Moment}
In principle, all moments of $N(t)$ can be calculated exactly,
following the same lines as in the previous Sec.~\ref{Sec:second-moment} and in App.~\ref{Sec:SecondMomentDerivation}. However,
even the second moment involves an integral that
cannot be carried out in closed form and, for
higher moments, the procedure quickly becomes
unpleasantly complicated.
To calculate higher moments of the population
size $\Expected{N^n(t)}$, we resort to a perturbative
expansion in powers of the amplitude $A$, which 
can be
systematically expressed in terms of diagrams shown in the appendix.

Following \cite{Garcia-Millan2018}, all moments can be written in terms of factorial moments, which
are naturally produced by the diagrammatics of the field theory. We denote the $n$-th factorial moment by $g_n(t_0,t)$,
so that
\begin{equation}\elabel{moments_in_factorial_moments}
    \Expected{N^n(t)|N(t_0)=1} = \sum_{\ell=0}^n 
    \begin{Bmatrix}
    n\\
    \ell
    \end{Bmatrix}
g_\ell(t_0,t) \ ,
\end{equation}
where $\begin{Bmatrix}n\\\ell\end{Bmatrix}$ are the Stirling numbers of the second kind. In the following, the factorial moments are expressed in orders of $A$,
\begin{equation}
    g_n(t_0,t)=g^{(0)}_{n}(t_0,t) + A g^{(1)}_{n}(t_0,t)+\OC(A^2).\label{eq:moment-approximation-notation}
\end{equation} 
The $n$-th factorial moment at $A=0$, given by
\begin{equation}
g^{(0)}_{n}(t_0,t)=n!e^{-r(t-t_0)}\left(\frac{q_2}{r}\left(1-e^{-r(t-t_0)}\right)\right)^{n-1},\label{eq:old-nth-factorial-moment}
\end{equation}
was 
calculated in closed form from the diagrams 
as a matter of combinatorics \cite{Garcia-Millan2018}. $g_n^{(0)}$ is dominated by $(q_2(t-t_0))^{n-1}$ for small $r(t-t_0)\ll1$, while it is exponentially decaying for large $r(t-t_0)\gg1$. 
In the present work, the factorial moments acquire a dependence on the initial time $t_0$, when
$N(t_0)=1$. Only to zeroth order in $A$, at $A=0$, do the factorial moments in the present work
become time-homogeneous and
reduce to those calculated in \cite{Garcia-Millan2018}, $g^{(0)}_{n}(t_0,t)=g^{(0)}_{n}(0,t-t_0)$.
The next order term in the small-$A$ expansion equals 
\begin{multline}
\label{eq:moments-first-order-correction}
    g^{(1)}_{n}(t_0,t)
= g_n^{(0)}(t_0,t) s \int_0^{t-t_0}  \sin(\nu (t-t'))\\
\times\left( 1+ (n-1)  \frac{e^{-rt'}-e^{-r(t-t_0)}}{1-e^{-r(t-t_0)}}\right) \dint t' \, ,
\end{multline}
whose derivation is explained in Appendix~\ref{Sec:NMomentDerivation}. 

In the subcritical regime $r>0$, in large $r(t-t_0)\gg1$ 
all moments vanish exponentially, because $g_n^{(0)}$ vanishes exponentially, see Eq.~\eqref{eq:old-nth-factorial-moment}. 
For small
$r (t-t_0)\ll 1$ and large $q_2(t-t_0)\gg 1$,
the moments are dominated by the largest factorial moment, 
\begin{equation}
   \text{for }rt\ll1:\qquad \lim_{t\rightarrow\infty}\frac{\Expected{N^n(t)}}{g_n(0,t)}=1.
\end{equation}
This can be seen by expanding $\mathbb{E}[N^n(t)]$ in terms of the factorial moments which are asymptotically dominated by  $g_n(t_0,t)\sim\OC\left((q_2t)^{n-1}\right)$,
\begin{multline}
\Expected{N^n(t)} = 
g_{n}(0,t) +\begin{Bmatrix}n\\2\end{Bmatrix}\underbrace{g_{n-1}(0,t)}_{\sim\OC\left((q_2t)^{n-2}\right)}+\dots
    \end{multline}
Within the small-$A$ expansion of the $n$-th factorial moment, the terms $g_n^{(0)}$ and $g_n^{(1)}$ dominate $\mathbb{E}[N^n(t)]$, such that  \begin{multline}\elabel{Nn_in_gn}
\Expected{N^n(t)} = 
g_{n}^{(0)}(0,t)+Ag_n^{(1)}(0,t)+\\+\OC\left(A^2(q_2t)^{n-1}\right) +\OC\left((q_2t)^{n-2}\right)
    \end{multline}   

However, from 
\Eref{moments-first-order-correction} it can be seen that the oscillations 
in the amplitude of $g^{(0)}_{n}(t_0,t)$ in $g^{(1)}_{n}(t_0,t)$ will never cease, so that the limit
$\lim\limits_{t\to\infty} \Expected{N^n(t)}/(q_2 (t-t_0))^{n-1}$ strictly does not exist. In other words,
$g_n(t_0,t)$ captures the leading order of $\Expected{N^n(t)}$ in $t$, but 
$(q_2 (t-t_0))^{n-1}$ does not.
 
The first two moments at $r=0$ can be approximated to first order in $A$ by 
\begin{subequations}
\elabel{gn_approx_of_moments}
\begin{align}\elabel{gn_approx_of_N1}
    \Expected{N(t)} & =
    1-\frac{As}{\nu}\left(
    \cos(\nu t) - 1 \right)  
    + \OC\!\left(A^2\right)\\
    \elabel{gn_approx_of_N2}
    \Expected{N^2(t)} & =
    2 q_2 t \left(
    1+ \frac{As}{\nu} 
    \left(
    1- 2 \cos(\nu t)\right) +\OC\!\left(A^2\right)\right)\nonumber\\
    &\quad 
    + \OC\!\left((q_2t)^0\right)
\end{align}
\end{subequations}
on the basis of \Eref{moments-first-order-correction} and \Eref{Nn_in_gn}. The expressions are consistent  with the exact expressions \Eref{exact-first-moment} and \Eref{exact-second-moment} (as given in \Eref{first-order-second-moment}) 
respectively. 
Figs.~\ref{fig:firstMoment} and \ref{fig:secondMoment} show a comparison between
the exact expressions (solid blue lines) and the
approximation \Eref{gn_approx_of_moments} (black dashed lines).
Deviations are clearly noticeable only for large amplitudes $A$. 

\section{Further Observables}
In the following, we analyse observables that 
are somewhat more involved to derive in the present
framework. In particular, the avalanche shape and covariance are 
more of immediate interest to experimentalists because these are more accessible from LFP recordings of the brain \cite{BeggsPlenz:2003,BeggsPlenz:2004,Priesemann:2009,Priesemann2013,Wilting2018,Wilting2018b,Wilting2019}.

\subsection{Covariance}
\label{Sec:Covariance}
The autocorrelation function 
\begin{equation}\elabel{def_covar}
\Cov(N(t_1),N(t_2))=
    \Expected{N(t_1)N(t_2)}
    - \Expected{N(t_1)}\Expected{N(t_2)}
\end{equation}
is the covariance 
of $N(t_1)$ and $N(t_2)$
and 
a common way to quantify how strongly correlated
data are at different points in time. 
If the population size at different times was independent, the autocorrelation function at times $t_1$ and $t_2$, $t_1\ne t_2$,
would be zero (the converse does not hold in general \cite{Park2018}).

As shown in App.~\ref{Sec:CovarianceDerivation} the covariance can
be calculated in closed form up to an integral,
\begin{align}\elabel{analytic_covar}
    &\Cov(N(t_1),N(t_2))
    =
    \Exp{-rt_\text{max}+\frac{As}{\nu}\left(\cos(\nu t_\text{max})-1\right)}\nonumber\\
    &\times \left[1+2q_2
    \int\limits_{0}^{t_\text{min}}\Exp{r(t'-t_\text{min})+\frac{As}{\nu}\left(\cos(\nu t_\text{min})-\cos(\nu t')\right)}\plaind t' \right]\nonumber\\
    &-\Exp{-r(t_1+t_2)+\frac{As}{\nu}\left(\cos(\nu t_1)+\cos(\nu t_2)-2\right)}\, ,
\end{align}
where $t_\text{min}=\min\{t_1,t_2\}$ and $t_\text{max}=\max\{t_1,t_2\}$.

\begin{figure}
    \centering
    \subfigure[\hspace{230pt}]{\label{fig:Cov1}\includegraphics[width=\columnwidth]{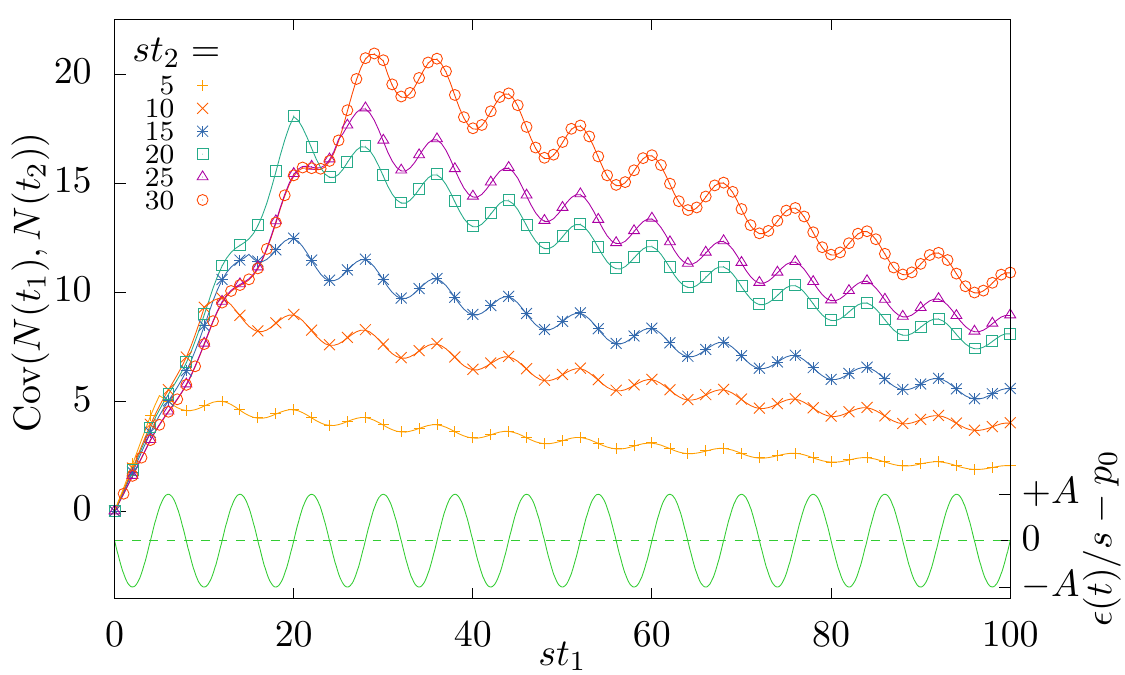}}
    \subfigure[\hspace{230pt}]{\label{fig:Cov2}\includegraphics[width=\columnwidth]{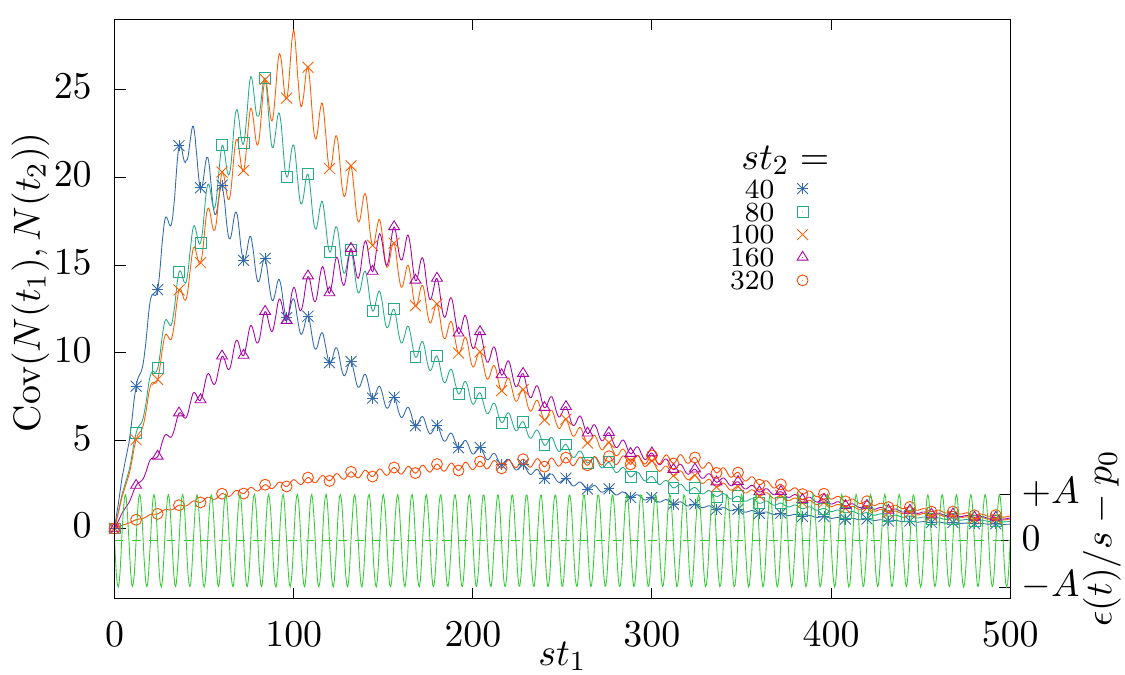}}
    \caption{Covariance $\Cov(N(t_1),N(t_2))$ \Eref{def_covar} for $r/s=0.01$, $\nu/s=\pi/4$, $A=0.05$ and
    \subref{fig:Cov1} $st_2=5,10,15,\ldots,30$,
     \subref{fig:Cov2} $st_2=40,80,100,160,320$.
    The system was initialized with a single particle at time $t_0=0$. Symbols: Simulations results from $10^9$ trajectories. Full lines: Analytic expression, \Eref{analytic_covar}.
    The perturbation of the extinction rate, \Eref{extinction-rate}, is shown in green at the bottom of the figure  (right ordinate).}
    \label{fig:covariance01}
\end{figure}

Simulation results and analytical expression \Eref{analytic_covar} are shown together in Fig.~\ref{fig:Cov1} and Fig.~\ref{fig:Cov2}. 
For $A=0$, the maximum of $\Cov(N(t_1),N(t_2))$ 
occurs
at $t_1=t_2=t$ 
and equals $(2 q_2/r + 1)(\exp{-rt}-\exp{-2rt})$ \cite{Garcia-Millan2018}. Furthermore, its maximum
rises with increasing $t$, provided that $rt<\log(2)$. When $A\neq0$, these results are only approximations and the exact position of the maximum depends on the phase and amplitude  of the oscillations. 
In \Fref{Cov1}, all $st_2$ are chosen to fulfil that condition, while in \Fref{Cov2}, the larger $st_2$ don't meet  this condition and show a decreasing maximum.

As an autocorrelation function, the covariance quantifies how the activity in the system at one instance influences the system at later instances. This has recently proven to be a valuable tool for determining the distance of neural networks from the critical point \cite{Wilting2018,Wilting2018b,Wilting2019}. What the imposed oscillations imply for this tool is discussed in Sec.~\ref{Sec:Conclusion}.

\subsection{Survival Probability}
\label{Sec:SurvProb}
The first moment, \Sref{first_moment}, shows that the expected number of particles is on average larger if $A>0$, i.e. if the extinction rate 
drops before growing in every period, compared to the case  without oscillations or reversed order of 
rise and fall of extinction, $A<0$. 
As noted in \Sref{first_moment}, 
the sign of the mass $r$ still determines whether $\Expected{N(t)}$ ultimately vanishes or diverges, even when
$\Expected{N(t)}$ oscillates indefinitely for $r=0$.
This observation raises the question whether 
the survival probability $P_s(t_0,t)$, that is 
the probability of $N(t)>0$ at a given time $t$ after initialization at $t_0$,
displays a corresponding behaviour.

Based on the derivation in App.~\ref{Sec:SurvProb_derivation}, 
to leading order in $A$
we find
\begin{align}\elabel{critical_Ps}
    \lim\limits_{r\rightarrow0}&P_s(t_0=0,t)=\,\frac{1}{1+q_2 t}\\
    &\,+\frac{As}{\nu}\left(\frac{1}{1+q_2t}-\frac{\cos(\nu t)+\frac{q_2}{\nu}\sin(\nu t)}{(1+q_2 t)^2}\right)+\mathcal{O}(A^2)\,.\notag
\end{align}

\begin{figure}
    \centering
    \includegraphics[width=\columnwidth]{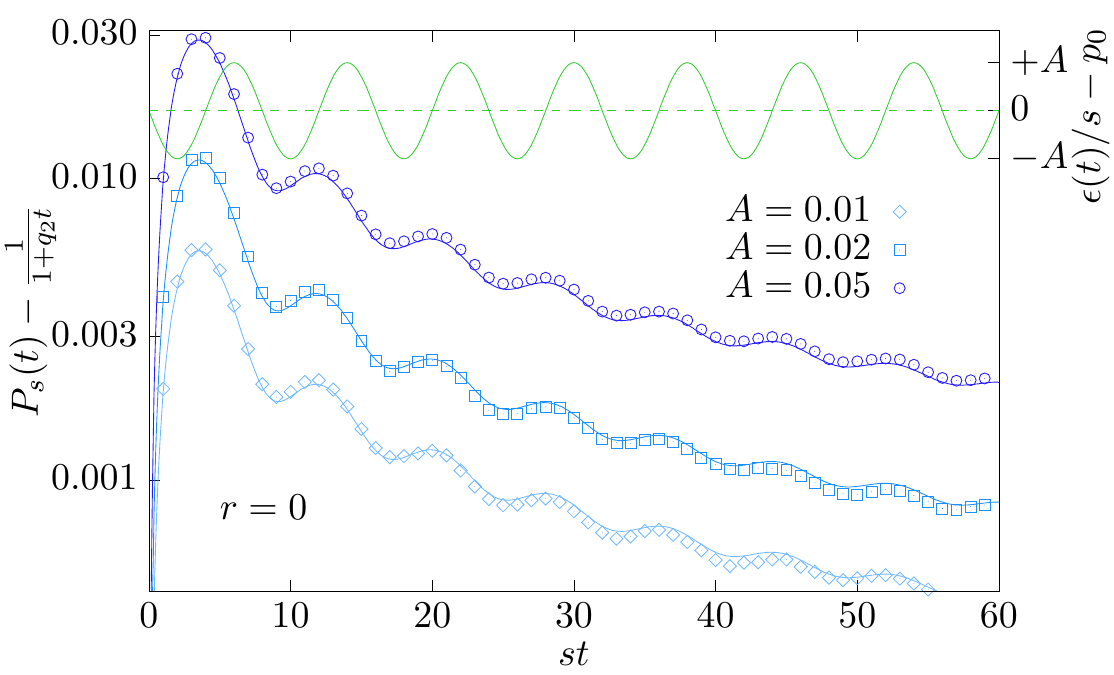}
    \caption{Difference between the probability of survival 
    $P_s(0,t)$ at $r=0$, \Eref{critical_Ps}
    with and without oscillations, for $\nu/s=\pi/4$ and $A\in\{0.01,0.02,0.05\}$. 
    The system was initialized with one particle at time $t_0=0$. Symbols: Simulations results using $3\cdot 10^7$ trajectories for $A=0.01$ and $10^7$ otherwise. Full blue lines: Analytic prediction to first order in $A$. 
    Full green lines: 
    Perturbation of the extinction rate, \Eref{extinction-rate}  (right ordinate).
    }
    \label{fig:prob-surv}
\end{figure}

The first term, which is independent of $A$, is the probability of survival of the critical branching process with constant extinction rate \cite{Garcia-Millan2018}, \Fref{prob-surv}. The second term is the first-order correction and indicates a shift of the probability of survival. For positive $A$ 
(leading to an initial decrease of the extinction rate)
the survival probability increases compared to the system without oscillations. 
For negative $A$ (corresponding to an initial increase of the extinction rate) it decreases. For $A>0$, the initial push into the supercritical phase seems to dominate the entire survival probability, even many oscillations later, see Fig.~\ref{fig:prob-surv}.

Despite the shifted survival probabilities, Fig.~\ref{fig:prob-surv}, and shifted average number of particles, Fig.~\ref{fig:firstMoment}, the avalanches do not survive indefinitely. They die eventually with probability $1$ at $r=0$. 
To appreciate better the effect of the extinction oscillations on the ultimate survival, we also consider
the asymptotics of large $t$ for $r>0$ or equivalently $p_0>p_2$,
\begin{multline}\elabel{survival_at_r>0}
\text{for }rt\gg1\\
P_s(0,t)|_{r>0} \simeq  \frac{\exp{-rt}}{1+q_2/r}
\left(
1+\frac{As}{\nu}(1-\cos(\nu t))
+\mathcal{O}(A^2)\right)\\
\xrightarrow{t\rightarrow\infty}0\hspace{5cm}
\end{multline}
to leading order.

For $r<0$, or equivalently $p_0<p_2$, the limit is positive,
\begin{equation}\elabel{active_phase}
\lim_{t\to\infty}
P_s(t_0,t)|_{r<0} = 
 -\frac{r}{q_2} \left(
 1 + \frac{As\nu}{r^2+\nu^2}
 \right)
 +\mathcal{O}(A^2),
\end{equation}
where we have to rely on \Eref{P_s_full} being the analytic continuation for the result obtained at positive mass $r$.
\Eref{survival_at_r>0} implies that $P_s(t_0,t)$, with or without oscillations, vanishes in the limit of large times as extinction prevails, since $r>0$. 
\Eref{active_phase} indicates that the linear increase in $-r$ of the ultimate survival probability is present
with or without oscillations, however, that 
the amplitude of 
that increase depends on $A$.
As far as the frequency $\nu$ is concerned, 
the effect of the oscillations on the ultimate survival 
is most pronounced for $\nu=\pm r$,
\begin{equation}\elabel{active_phase_special_nu}
\lim_{t\to\infty}
P_s(t_0,t)|_{r<0} = 
 -\frac{r}{q_2} \pm \frac{As}{2q_2}
+\mathcal{O}(A^2),
\end{equation}
with the minimum attained if $\nu=r$ and the maximum for $\nu=-r$.
It is noteworthy that \Eref{active_phase_special_nu} no longer vanishes as $r\to0$. 
The constraints mentioned above, such as $As\le r+q_2$, do not affect this result, as
$As=q_2/2$ still produces 
$\lim\limits_{r\to0^-}
\lim\limits_{t\to\infty}
P_s(t_0,t)|_{r<0} = 1/4$.
Together with \Eref{critical_Ps} this seems to suggest the possibility of a 
sudden onset survival, whereby 
$\lim\limits_{t\to\infty} P_s(t_0,t)$ 
jumps from $0$ at $r\to0$ to a finite value. 

However, it is crucial in which order limits are taken. \Eref{critical_Ps} remains valid if $r\to0$ (tying $\nu=\pm r$) \textit{before} $t\to\infty$. When taken in this order, the ultimate survival probability is zero at the critical point.

We therefore conclude that the ultimate survival increases linearly and continuously from $0$ for $r\ge 0$, \Erefs{critical_Ps} and \eref{survival_at_r>0}, to 
a finite value at $r<0$, \Eref{active_phase}. If the critical point of the present process is defined as the onset of ultimate survival, 
then it is not shifted by the oscillations. 
Fig.~\ref{fig:prob-surv} shows a comparison of the first-order corrected survival probability \Eref{critical_Ps}
as a function of time to simulation results.

\subsection{Avalanche Shape}
\label{Sec:Shape}

\subsubsection{Shape depending on time of death T}
\begin{figure}
    \centering
    \includegraphics[width=\columnwidth]{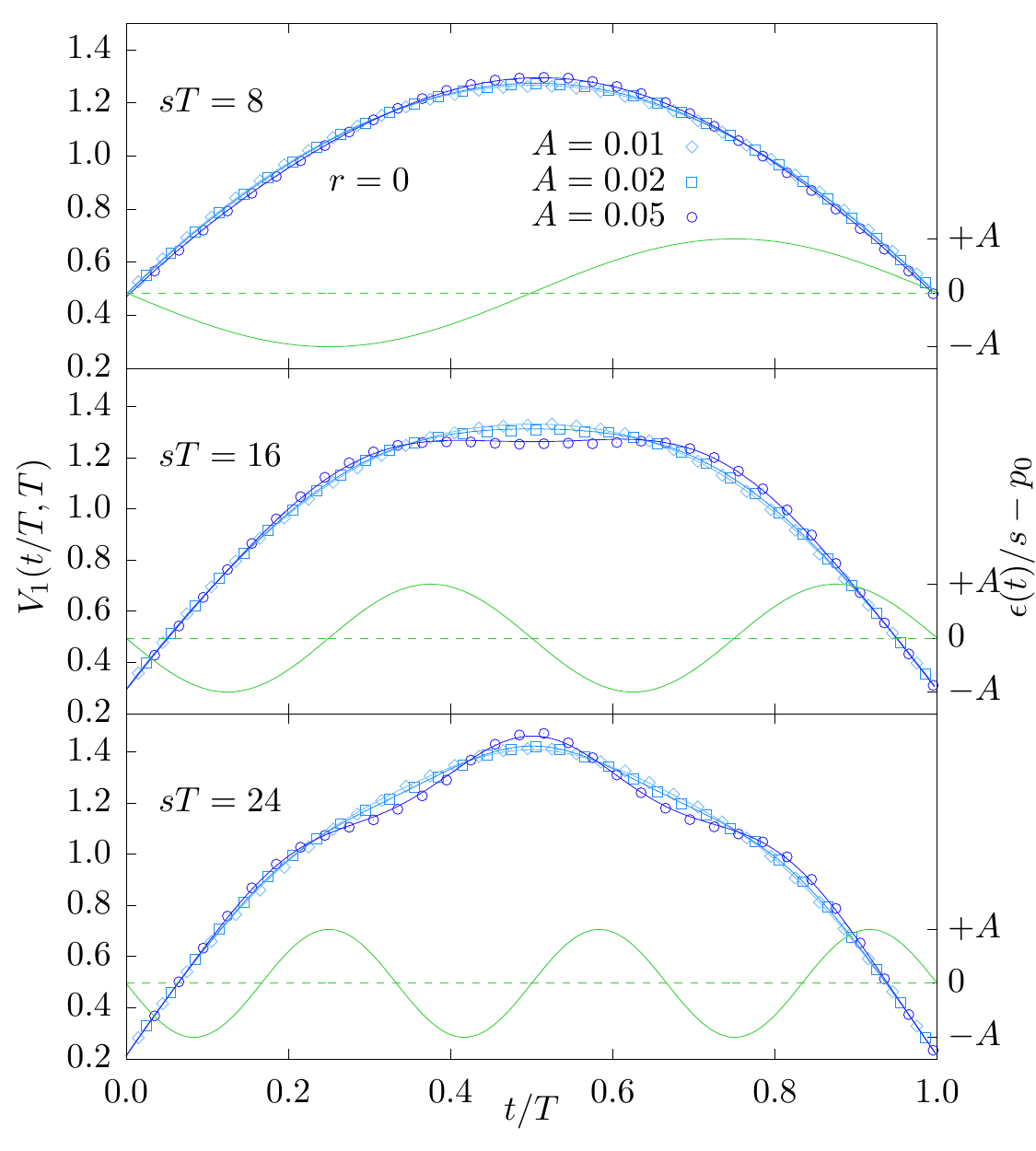}
    \caption{Area-normalised expected avalanche shapes 
    $V_1(t,T)$, \Eref{def_V_1}, for $r=0$, $\nu/s=\pi/4$ and $sT\in\{8,16,24\}$ for $A\in\{0.01,0.02,0.05\}$. The time is rescaled by the termination time $T$. The system was initialized with one particle at time $t_0=0$. Symbols: Simulation results obtained by averaging over $10^8$ trajectories $N(t)$ with a termination time $sT\pm0.2$. Full blue lines: Analytic prediction to first order in $A$.
    Full green lines: Perturbation of the extinction rate, \Eref{extinction-rate}  (right ordinate).}
    \label{fig:shape01}
\end{figure}

\newcommand{\flabel}[1]{\label{fig:#1}}

\begin{figure}
    \centering
    \includegraphics[width=\columnwidth]{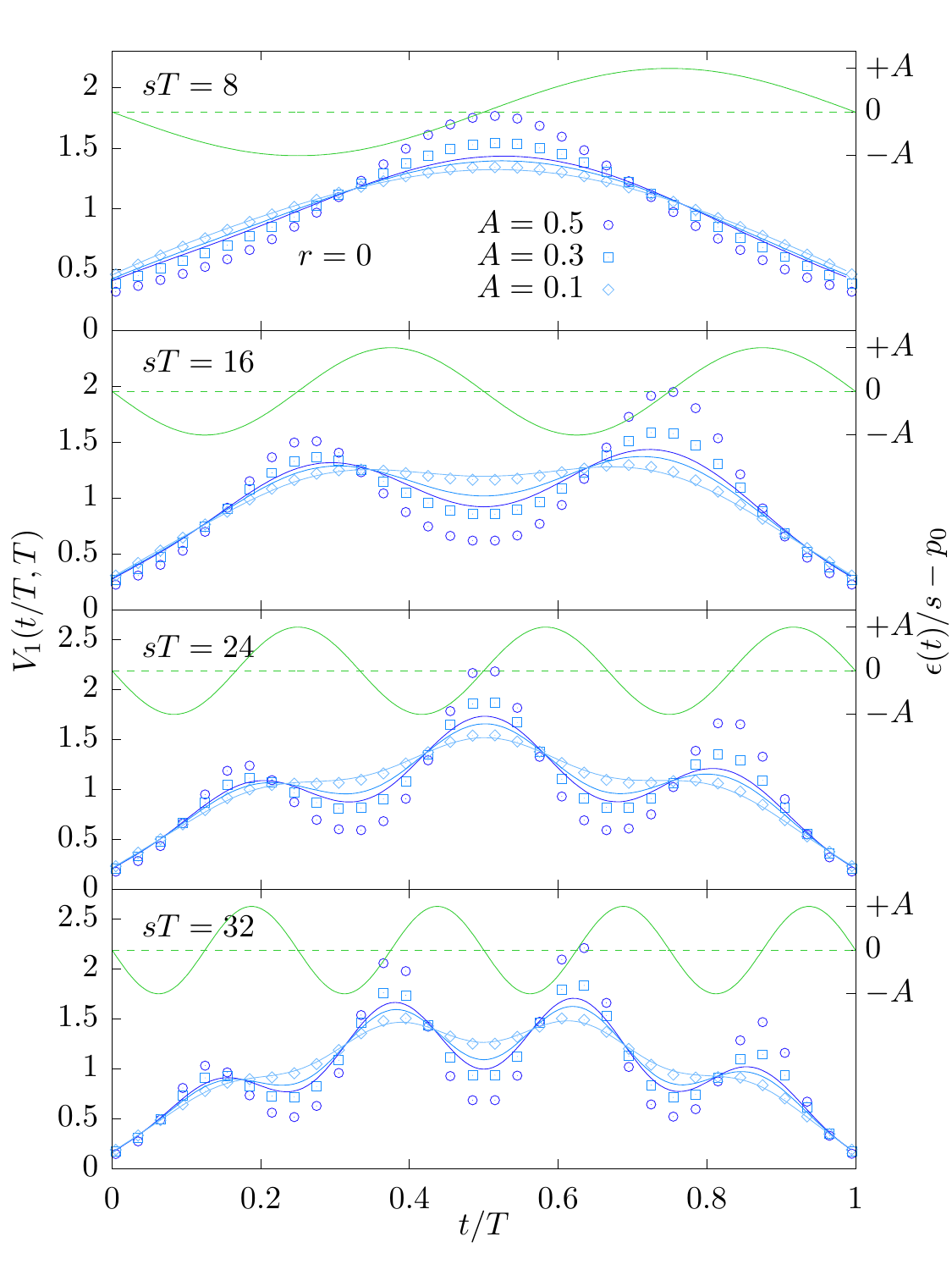}
    \caption{
    Area-normalised expected avalanche shapes 
    $V_1(t,T)$, \Eref{def_V_1},
    for $r=0$, $\nu/s=\pi/4$ and $sT\in\{8,16,24,32\}$ for $A\in\{0.1,0.3,0.5\}$. The time is rescaled by the termination time $T$. The system was initialized with one particle at time $t_0=0$. Symbols: Simulations results obtained by averaging over $10^8$ trajectories $N(t)$ with a termination time $sT\pm0.2$. Full blue lines: Analytic prediction to first order in $A$.
    Full green lines: Perturbation of the extinction rate, \Eref{extinction-rate}  (right ordinate).}
    \flabel{largeA_shape}
\end{figure}

The (temporal) shape of the avalanche  
$V(t,T)$
is the expectation of $N(t)$ conditioned to the 
branching process going spontaneously
extinct at some fixed termination time $t=T$, \ie for all small $\delta>0$, $N(T-\delta)>0$ and $N(T+\delta)=0$. Typically, time is rescaled to $\tau=t/T$ \cite{KuntzSethna:2000}.

After suitable normalisation \cite{Garcia-Millan2018} the time-homogeneous
version of the process processes displays a universal, parabolic shape. 
Not least because of its universality, the shape has gained some popularity 
to serve as a fingerprint of a process \cite{DobrinevskiLeDoussalWiese:2014,Papanikolaou2011,BaldassarriColaioriCastellano:2003}.

As shown in Figs.~\ref{fig:shape01} and \ref{fig:largeA_shape}, the periodic extinction imposes characteristic humps on the shape.
They are of course rooted in the periodic extinction, as an analytical calculation shows,
App.~\ref{Sec:ShapeDerivation}.
These oscillations remain visible in the shape $V(t,T)$ at all $T$ and all $t$, but
become particularly vivid whenever the termination time $T$ is commensurate with the period of the oscillations,
$2\pi/\nu$. As we have used our field-theoretic scheme only to first order, the slight mismatch with simulation results at larger $A$, such as those shown in \Fref{largeA_shape}, is not surprising.
However, at such large amplitudes, the resulting shape resembles that of recent experimental results, where $\gamma$-oscillation modulated the average shape of neuronal avalanches \cite{MillerYuPlenz:2019}. Furthermore, for larger amplitudes $A$, the avalanche shapes are asymmtric, which can be seen particularly well in Fig.~\ref{fig:largeA_shape} for $sT=16$.
\subsubsection{Comparison to Experiments}
\begin{figure*}
    \centering
    \includegraphics[width=\textwidth]{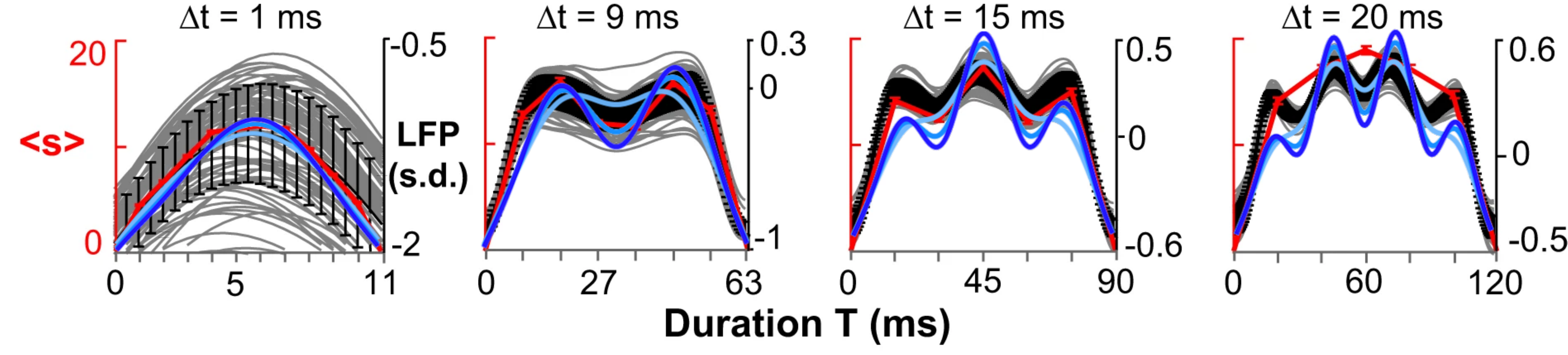}
    \caption{Comparison of model to experimental data. Underlying figure (grey, black and red colors) from Miller, Yu and Plenz (2019) \cite{MillerYuPlenz:2019}, Fig.~4(a), reproduced and adapted under Creative Commons Attribution 4.0 International License. Adaptation: Overlay of blue colors: analytical results from Fig.~\ref{fig:largeA_shape} above. On $y$-axis, $\langle s\rangle$ is the mean profile in \cite{MillerYuPlenz:2019}, corresponding to our $V(t,T)$; on $x$-axis, duration T from \cite{MillerYuPlenz:2019} is denoted by time $t$ in our article. Data was collected through multielectrode arrays implanted in adult nonhuman primates (see \cite{MillerYuPlenz:2019} for details). Grey lines, single electrode data, black line mean of array, red mean size-per-timestep. }
    \flabel{shape-experiment}
\end{figure*}

The mean avalanche shapes, such as in Figs.~\ref{fig:shape01} and \ref{fig:largeA_shape} can be qualitatively compared to avalanche profiles recorded in the brain. We reproduce in Fig.~\ref{fig:shape-experiment} plots from Miller, Yu and Plenz (2019)  (Figure 4(a) in reference \cite{MillerYuPlenz:2019}), and adapt it by overlaying our analytical plots from Fig.~\ref{fig:largeA_shape}, in accordance with the Creative Commons Attribution 4.0 International License. The data was collected through multielectrode arrays implanted in three adult nonhuman primates (see \cite{MillerYuPlenz:2019} for details).

The plots show qualitatively a good agreement between the data and our model. The general shape and periodicity is captured extremely well. Small qualitative disagreement occurs in the third and fourth plot in the relative heights of consecutive maxima and minima.
This disagreement may be down to higher order 
correction terms in the amplitude $A$.

The figure illustrates that branching processes, which are commonly used to explain the statistical properties of avalanches recorded in the brain, can be extended to also incorporate neuronal oscillations. Although the observation of non-universal avalanche shapes questioned the criticality hypothesis of the brain\cite{Beggs:2008,Wilting2019b,MillerYuPlenz:2019}, our analytical results clearly show that criticality is compatible with avalanche profiles that are modulated by oscillations.  

\subsubsection{Universal Parabolic Shape}
\begin{figure}
    \centering
    \includegraphics[width=\columnwidth]{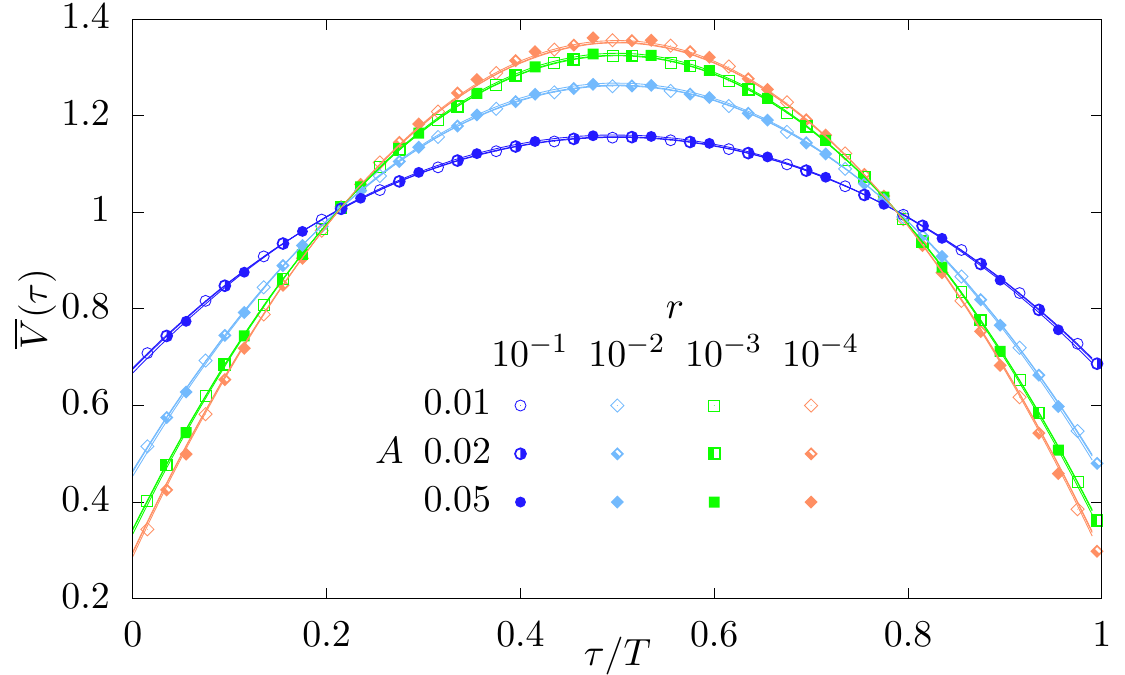}
    \caption{Area-normalized expected avalanche shapes averaged over time of death $T$,
    $\overline{V}(\tau)$, \Eref{def_ave_V},
    for $r\in\{10^{-4},10^{-3},10^{-2},10^{-1}\}$, $\nu/s=\pi/4$ and $A\in\{0.01,0.02,0.05\}$. Before averaging, the time of each avalanche is rescaled by its termination time $T$. The system was initialized with one particle at time $t_0=0$. Symbols: Simulations results obtained by averaging over $10^7$ trajectories $N(t)$. Full lines: Analytic prediction to first order in $A$.}
    \flabel{ave_shape}
\end{figure}
The universal parabolic shape 
may be recovered by suitably averaging across different termination times $T$. 
Devising such a scheme in a given numerical or experimental setting may not always be feasible \cite{WillisPruessner:2018}. 
\Fref{ave_shape} shows the $T$-averaged and $T$-rescaled expected avalanche shape,
\begin{equation}\elabel{def_ave_V}
    \overline{V}(\tau) = \frac{\int_0^\infty P_T(T) V(\tau T,T) \plaind T }
    {\int_0^1\int_0^\infty P_T(T) V(\tau T,T) \plaind{T}\plaind\tau}.
\end{equation}
Varying small amplitudes do not seem to alter this shape and appear to converge to the universal parabola shape when criticality is approached $r\rightarrow0$.

\section{Discussion and Conclusion}
\label{Sec:Conclusion}
Our discussion is two-fold: First, we focus on aspects of interest to research in stochastic processes. We then follow with a discussion on the implications for research in the area of neuronal avalanches. 

In this paper, we extend the standard branching process by including time-dependent, deterministic variations of a reaction rate. While we focus on varying the extinction rate only, our approach is applicable to any reaction rate in a Doi-Peliti field theory. 

Neuronal avalanches can show oscillating behaviour, which we capture with a  branching process model with oscillating extinction rate. These oscillations can be observed in any moment, of which we present the zeroth (the survival probability), first and second, and in any correlation function, of which we present the two-time covariance. All of these observables are calculated exactly. Furthermore, we introduce an approximation scheme for factorial moments to first order in the amplitude of the extinction rate oscillations. This allows approximating more complicated observables such as the survival probability and the avalanche shape in the limit of small oscillation amplitudes. 

All analytically calculated observables are compared with  simulation results from Monte Carlo simulations. While the exact analytic results match perfectly, we also evaluate the first-order approximation scheme and find good agreement in the limit of small oscillation amplitudes. 

Although the extinction rate is unchanged on average, its oscillation leads \textit{on average} to a shift in the moments. For example, if the oscillations start with a decrease of the extinction rate, the expected particle number is always greater compared to the process without oscillations. Conversely, if the oscillations start by increasing the extinction rate, the expected number of particles is always lower than without oscillations. Despite the average shift of particle numbers, the onset of indefinite survival is not shifted by the oscillations. We therefore conclude that the critical point remains unchanged. 

Both subcritical branching processes and neuronal activity show exponentially decaying auto-correlation functions whose decay rate is independent of the spatial sub-sampling of the neural network \cite{Wilting2019}. In particular, if the neural system is close to the critical point, in the reverberating regime, the exponential decay 
is slow and
can be observed over more than 100ms even in a single neuron's 
activity \cite{Wilting2019}. As the oscillations in the neuronal activity can have periods well below 100ms (\eg in the spindle band), they should be visible in the data. 

Similarly, both the oscillating branching process presented here and recordings of neuronal avalanches show oscillating avalanche shapes \cite{MillerYuPlenz:2019}. In our model and in the data, the oscillations modulate the shape and are most pronounced when the avalanche length is an integer multiple of the period of the oscillations. Surprisingly, these results defy the assumption that the avalanche shape is universal at criticality. However, when the shapes are also rescaled and averaged over the avalanche duration, our simulations indicate that the universal parabola shape is recovered.

In future research, the oscillating branching process should be compared quantitatively to local field potential recordings of oscillating neuronal avalanches. Thus, the presented model can contribute to the understanding of neuronal avalanches by rejecting or supporting the branching process picture of  signal propagation in the brain. Furthermore, it would be interesting to find out whether other types of variation of the extinctions rate, such as random fluctuations, result in similar behaviour.

\section*{Acknowledgements}
We would like to thank Stephanie Miller for fruitful discussion
and for sharing unpublished work. We would also like to thank Dietmar Plenz for valuable feedback. R.G.M.~is grateful to the Santa
Fe Institute, the Imperial College Dean's Fund, the Doris Chen 
Mobility award and G.P.~for their support to attend the SFI Complex Systems
Summer School 2018 in New Mexico, where she met Stephanie Miller.
J.P.~was partially supported by an EPSRC Doctoral Prize Fellowship.
All authors have benefited from Andy Thomas' tireless computing support.

\section*{Author Contributions}
All authors contributed equally.
\section*{Competing Interests Statement}
The authors declare no competing interests.

\appendix

\section{Extension of the Action}
\label{Sec:DerivationAction}
The extension of the model consists of making the extinction rate time-dependent. Hence, in the following derivation, we focus on introducing the new extinction rate and ignore the branching process. 

Let $P(n,t)$ be the probability that there are $n$ particles in the system at time $t$, \ie $N(t)=n$.  Then, the master equation for the extinction process only is\begin{align}\label{Eq:masterEq}
    \frac{\plaind}{\plaind t}P(n,t)=\epsilon(t)\big((n+1)P(n+1,t)-nP(n,t)\big).
\end{align}
The first step in the derivation of the field theory is the introduction of \textit{bra-ket vectors} $\langle n|$ and $|n\rangle$, respectively, and \textit{ladder operators} $a$ and $a^\dagger$. The ket vector $|n\rangle$ represents that there are $n$ particles in the system, while the bra vector $\langle n|$ project out the state of $n$ particles, $\langle n | m \rangle=\delta_{n,m}$, the Kronecker function. The ladder operator $a^\dagger$ creates an additional particle, \ie $a^\dagger|n\rangle=|n+1\rangle$, and the operator $a$ annihilates a particle (with the additional prefactor $n$) $a|n\rangle=n|n-1\rangle$. Thus, the entire, probabilistic state of the system at time $t$ can be represented by a ket vector $|\mathcal{M}(t)\rangle$, \begin{align}
    |\mathcal{M}(t)\rangle=\sum\limits_{n=0}^\infty P(n,t)|n\rangle,
\end{align}
which is nothing else than a representation of the probability generating function. 

In the next step, following \cite{Doi:1976}, the master equation~\eqref{Eq:masterEq} is used to derive a corresponding equation for the state vector $|\mathcal{M}(t)\rangle$,
\begin{align}\label{Eq:SQEQ}
    \frac{\plaind}{\plaind t}\Big|\mathcal{M}(t)\Big\rangle=\epsilon(t)\left(a-a^\dagger a\right)\big|\mathcal{M}(t)\big\rangle.
\end{align}

In the final step, based on work by Peliti \cite{Peliti:1985}, equations such as~\eqref{Eq:SQEQ} can be solved by a path integral with an action $\mathcal{A}_\epsilon$, where in every normal-ordered compound operator, $a$ is replaced by a field $\phi(t)$ and $a^\dagger$ is replaced by a field $\phi^\dagger(t)$, and where an additional time-derivative term is added:
\begin{align}
    \mathcal{A}_\epsilon=\int\!\! \big(\phi^\dagger(t)-1\big)\left(-\frac{\plaind}{\plaind t}\phi(t)\right)+\epsilon(t)\big(\phi(t)-\phi^\dagger(t)\phi(t)\big)\plaind t
    \, .
\end{align}
In general, it is useful to shift the creation field $\phi^\dagger=\widetilde\phi+1$. An explanation of implications of the shift, as well as a more detailed derivation of the path integral can be found in \cite{Pausch2019}. Thus, a simpler action is obtained \begin{align}
    \mathcal{A}_\epsilon=\int \widetilde\phi\left(-\frac{\plaind}{\plaind t}-\epsilon(t)\right)\phi(t)\plaind t \ .
\end{align}
The total action 
\begin{equation}\elabel{action}
    \mathcal{A} = 
    \int \left\{
    \widetilde\phi\left(-\frac{\plaind}{\plaind t}-\epsilon(t)\right)\phi(t)
    +\sum\limits_{j=2}^\infty q_j\widetilde\phi^j\phi
    \right\}\plaind t
\end{equation}
enters the path integral in the form $\exp{\mathcal{A}}$
so that field theoretic expectations become
\begin{equation}
    \ave{\bullet} = 
    \int  \Exp{\mathcal{A}} \bullet
    \mathcal{D} \phi \mathcal{D} \widetilde\phi
\end{equation}
and are calculated by performing the path integral over the bilinear part of the
action $\mathcal{A}_0$, \Eref{birth-death-action}, and expanding perturbatively 
for the term 
with coupling $sp_0-\epsilon(t)$, which is
unaccounted 
for in $\mathcal{A}_0$.
In connection with the observation of oscillating neuronal avalanches, we choose $\epsilon(t)=sp_0-As\sin(\nu t)$
and the perturbative part of the action becomes
\Eref{A-term}, giving rise to a perturbation theory in small $A$.

\section{Derivation of moments}
In order to keep a readable, consistent notation,
we use the mathematical notation, \eg $\mathbb{E}[N(t)]$,
when we
characterize the statistical properties of $N(t)$. 
When referring to field-theoretic calculations, we use the notation common in physics, \ie $\ave{\mathcal{O}}$ for the expectation of an observable $\mathcal{O}$. 
Of course, the two are
closely related, \eg $\ave{\phi(t)\phi^\dagger(t_0)}=\mathbb{E}[N(t)|N(t_0)=1]$
but
$\ave{\phi^2(t)\phi^\dagger(t_0)}+
\ave{\phi(t)\phi^\dagger(t_0)}
=\mathbb{E}[N^2(t)|N(t_0)=1]$,
\Eref{2nd_moment}.
Unless stated otherwise, observables are conditioned to the initialization of the branching process with one particle at time $t_0=0$, \ie $N(0)=1$,
routinely omitted in the mathematical notation. For example,
$\mathbb{E}[N(t)]$ is short for
$\mathbb{E}[N(t)|N(0)=1]$.

\subsection{First Moment}
\label{Sec:FirstMomentDerivation}
In the field theoretic calculation, the first moment is calculated as the propagator. Due to the varying extinction rate the propagator 
acquires corrections which schematically depicted as diagrams. The red, full lines 
\tikz[baseline=-2.5pt,scale=0.8]{
  \node at (1,0) [above=0.1pt] {$t_1$};
  \node at (0,0) [above=0.1pt] {$t_2$};
\draw[Aactivity]  (0,0) -- (1,0);
}
represent the bare propagator $\langle{\phi(t_2)\tildephi(t_1)}\rangle_0$, while the black dashed lines ending in circles 
\tikz[baseline=-2.5pt,scale=0.8]{
  \node at (0,0) [above=0.1pt] {$t'$};
  \node (a) at (0.75,0) {};
\draw[density]  (0,0) -- (a);
\draw[very thick,fill=white] (a) circle (2pt);
} represent correction terms, where $t_2\geq t'\geq t_1$. In a Doi-Peliti field theory, these diagrams are by convention read from right to left.
The first moment is thus written as follows: 
\begin{subequations}
\begin{align}
    &\mathbb{E}[N(t)|N(t_0)=1] 
    =\,\left\langle\phi(t)\phi^\dagger(t_0)\right\rangle
    =\ave{\phi(t)\phitilde(t_0)} \nonumber\\
    &\,\hat=\,
    \tikz[baseline=-2.5pt]{\draw[Aactivity](0,0) -- (1,0);}
    +
    \tikz[baseline=-2.5pt]{
    \draw[Aactivity](0,0) -- coordinate[midway] (m) (1,0);
    \draw[density]  (m) -- ++(-30:0.5);
    \draw[very thick,fill=white] (m)+(-30:0.5) circle (2pt);
    }
    +
    \tikz[baseline=-2.5pt]{
    \node (a) at (0,0) {};
    \node (b) at (1.8,0) {};
    \node (m1) at ( $ (a)!0.25!(b) $ ) {};
    \node (m2) at ( $ (a)!0.55!(b) $ ) {};
    \draw[Aactivity](a) -- (b);
    \draw[density]  (m1) -- ++(-30:0.5);
    \draw[very thick,fill=white] (m1)+(-30:0.5) circle (2pt);
    \draw[density]  (m2) -- ++(-30:0.5);
    \draw[very thick,fill=white] (m2)+(-30:0.5) circle (2pt);
    }
    +\dots\\
    &=\,\Theta(t)e^{-rt}\biggl(1+\int\limits_{t_0}^tsA\sin(\nu t')\plaind t'+\notag\\
    &\quad+\int\limits_{t_0}^tsA\sin(\nu t')\int\limits_{t_0}^{t'}sA\sin(\nu t'')\plaind t''\plaind t'+\dots\biggr)\notag\\
    &=\,\Theta(t)e^{-rt}\left(\sum\limits_{\ell=0}^\infty\frac{1}{\ell!}\int\limits_{t_0}^t sA\sin(\nu t')\plaind t'\right)\notag\\
    &=\,\Theta(t)\text{exp}\left(-
    r+\int\limits_{t_0}^tsA\sin(\nu t')\plaind t'\right),
    \elabel{first_moment_general}
\end{align}
\end{subequations}
exactly as expected for an inhomogeneous Poisson process with time-varying intensity that governs the extinction \cite{Kingman1992}. For $t_0=0$, this reproduces the result for
$\Expected{N(t)}$, \Eref{exact-first-moment}.

\subsection{Second Moment}
\label{Sec:SecondMomentDerivation}
The second moment $\mathbb{E}[N^2(t)]$ is determined field-theoretically
on the basis of the square of the particle number operator, $(a^\dagger a)^2=\left(a^\dagger\right)^2a^2 + a^\dagger a$, which
is easily expressed in terms of fields $\phi^\dagger$ and $\phi$ once 
in normal-ordered form,
\begin{equation}
\elabel{2nd_moment}
    \mathbb{E}[N^2(t)]=\langle\phi^2(t)\phitilde(0)\rangle+\underbrace{\langle\phi(t)\phitilde(0)\rangle}_{=\mathbb{E}[N(t)]}.
\end{equation}
Diagrammatically, the first term is 
\begin{align}
    \langle\phi^2(t)\phitilde(0)\rangle\hat=\,
    \tikz[baseline=-2.5pt]{\draw[very thick, color=red] (-0.5,0.4) -- (0,0) -- (0.5,0.0);\draw[very thick, color=red](-0.5,-0.4) -- (0,0);}\label{eq:second-moment-derivation}
\end{align}
which is a convolution in direct time,
\begin{align}
\langle\phi^2(t)\phitilde(0)\rangle =
2q_2\int\limits_{0}^t
\ave{\phi(t)\phitilde(t')}^2 \ave{\phi(t')\phitilde(0)}
\plaind t'\ ,
\elabel{secondmom_from_firstmom}
\end{align}
that can be calculated with the help of \Eref{first_moment_general}
to produce \Eref{exact-second-moment}.

\subsection{n-th factorial moment to first order}
\label{Sec:NMomentDerivation}
We define the $n$-th factorial moment of $N(t)$ \cite{Garcia-Millan2018} as the function 
 \begin{subequations}
\begin{align}
& g_n(t_0,t) = \ave{\phi(t)^n \tildephi(t_0)}\\
&= g_n^{(0)}(t_0,t) + A g_n^{(1)}(t_0,t) + A^2g_n^{(2)}(t_0,t) +\ldots
\end{align}
\end{subequations}
where the
additional extinction 
$-A\sin(\nu t)$,
\Eref{extinction-rate},
is dealt with perturbatively about $A=0$.
The zeroth-order correction $g_n^{(0)}(t_0,t)$ corresponds to the sum of binary tree diagrams with 
one in-coming leg and $n$ out-going legs. 

Since the zeroth-order correction does not include any perturbation in the extinction rate,
the process is homogeneous in time and, therefore,
it does not depend on the initial time $t_0$, but rather on the amount of time $t-t_0$.

In \cite{Garcia-Millan2018}, the function $g_n^{(0)}(t_0,t)$ was 
derived from the following forest (meaning it contains a sum of tree-like diagrams)
\begin{subequations}
\elabel{all_gn0}
\begin{align}
 g_n^{(0)}(t_0,t) 
&\corresponding\!\!
\sum_{\ell=1}^n \sum_{m_1,\ldots,m_\ell=1}\!\!\binom{n}{m_1,\ldots,m_\ell}\!\!
\tikz[baseline=-2.5pt,scale=0.8]{
\draw[Aactivity] (0.5,0) -- (0,0);
  \path [postaction={decorate,decoration={raise=0ex,text along path, text align={center}, text={|\large|......}}}] (170:0.7cm) arc (170:220:0.8cm);
  \node at (190:0.9) {$\ell$};
  \node (a) at (170:1.5) {};
  \path [postaction={decorate,decoration={raise=0ex,text along path, text align={center}, text={|\large|....}}}] (a)+(170:0.7cm) arc (170:210:0.8cm);
  \draw[Aactivity] (a)+(-150:0.2) -- ++(-150:0.8);
  \draw[Aactivity] (a)+(170:0.2) -- ++(170:0.8);
  \draw[Aactivity] (a)+(150:0.2) -- ++(150:0.8);
\draw[Aactivity] (a)+(0:0.2) -- (0,0);
\draw[thick,fill=white] (a)+(0,0) circle (0.2cm);
\node at (178:2.5) {$m_2$};
\begin{scope}
  \node (b) at (140:1.75) {};
  \path [postaction={decorate,decoration={raise=0ex,text along path, text align={center}, text={|\large|....}}}] (b)+(170:0.7cm) arc (170:210:0.8cm);
  \draw[Aactivity] (b)+(-150:0.2) -- ++(-150:0.8);
  \draw[Aactivity] (b)+(170:0.2) -- ++(170:0.8);
  \draw[Aactivity] (b)+(150:0.2) -- ++(150:0.8);
\end{scope}
\draw[Aactivity] (b)+(0:0.2) -- (0,0);
\draw[thick,fill=white] (b)+(0,0) circle (0.2cm);
\node at (160:2.5) {$m_1$};
\begin{scope}
  \node (c) at (-140:1.75) {};
  \path [postaction={decorate,decoration={raise=0ex,text along path, text align={center}, text={|\large|....}}}] (c)+(170:0.7cm) arc (170:210:0.8cm);
  \draw[Aactivity] (c)+(-150:0.2) -- ++(-150:0.8);
  \draw[Aactivity] (c)+(170:0.2) -- ++(170:0.8);
  \draw[Aactivity] (c)+(150:0.2) -- ++(150:0.8);
\end{scope}
\draw[Aactivity] (c)+(0:0.2) -- (0,0);
\draw[thick,fill=white] (c)+(0,0) circle (0.2cm);
\node at (207:2.6) {$m_\ell$};
}\\
&\corresponding \, n! \exp{-r(t-t_0)} \left( \frac{q_2}{r} \left( 1-\exp{-r(t-t_0)}\right)\right)^{n-1} \, .
 \elabel{gn0}
\end{align}
\end{subequations}
This zeroth-order correction satisfies the identity
\begin{widetext}
\begin{align}
g_n^{(0)}(t_0,t) 
& = \sum_{\ell=1}^n g_\ell^{(0)}(t',t) \frac{1}{\ell!} \sum_{m_1=1, \ldots, m_\ell=1}^n
\binom{n}{m_1, \ldots, m_\ell} g_{m_1}^{(0)}(t_0,t') \cdots g_{m_\ell}^{(0)}(t_0,t') \,, 
\elabel{gn0t2}
\end{align}
\end{widetext}
which can be obtained by choosing an arbitrary time $t'\in[t_0,t]$ when 
a diagram has $\ell$ legs, \cf 
\Eref{cut_tree}, and reconstructing the entire forest.
As the left-hand side of \Eref{gn0t2} does not depend on $t'$ 
it might be tempting to differentiate with respect to $t'$, 
but for our purposes the sum
\begin{widetext}
\begin{equation}\elabel{gn_time_derivative}
    \frac{\partial}{\partial t} g_n^{(0)}(t_0,t)
 = \sum_{\ell=1}^n \frac{\partial}{\partial t}g_\ell^{(0)}(t',t) \frac{1}{\ell!} \sum_{m_1=1, \ldots, m_\ell=1}^n
\binom{n}{m_1, \ldots, m_\ell} g_{m_1}^{(0)}(t_0,t') \cdots g_{m_\ell}^{(0)}(t_0,t')
\end{equation}
\end{widetext}
will prove particularly useful.

We proceed with the construction of the first order correction
$A g^{(1)}_n(t)$ of $g_n(t)$.
This correction  
contains the external perturbation $\plaind t' As\sin(\nu t')$
once in each diagram. It needs to be attached to every tree diagram at time $t'$,
as indicated by the vertical dotted line in the diagram below,\begin{widetext}
\begin{subequations}
\begin{align}
g_n^{(1)}(t_0,t)&\,\corresponding
\int_{t_0}^t
\sum_{\ell=1}^n 
\sum_{m_1,\ldots,m_\ell}^n
\!\!\binom{n}{m_1,\ldots,m_\ell}\!\!
\tikz[baseline=-2.5pt,scale=0.8]{
\node (c) at (-140:1.75) {};
\node (m) at ( $ (c)!0.45!(0,0) $ ) {};
\node (s) at (0.3,-1) {};
\draw[very thick,fill=white] (s) circle (2pt);
\draw[black, dotted, thick] ( $ (m)+(0.1,-0.7) $ ) -- ( $ (m)+(0.1,1.7) $ ) node [at start, right] {$\!t'$};
\draw[Aactivity] (0.5,0) -- (0,0);
  \path [postaction={decorate,decoration={raise=0ex,text along path, text align={center}, text={|\large|.......}}}] (170:1.2cm) arc (170:220:1.2cm);
  \node at (190:1.35) {$\ell$};
  \node (a) at (170:1.5) {};
  \path [postaction={decorate,decoration={raise=0ex,text along path, text align={center}, text={|\large|....}}}] (a)+(170:0.7cm) arc (170:210:0.8cm);
  \draw[Aactivity] (a)+(-150:0.2) -- ++(-150:0.8);
  \draw[Aactivity] (a)+(170:0.2) -- ++(170:0.8);
  \draw[Aactivity] (a)+(150:0.2) -- ++(150:0.8);
\draw[Aactivity] (a)+(0:0.2) -- (0,0);
\draw[thick,fill=white] (a)+(0,0) circle (0.2cm);
\node at (178:2.5) {$m_2$};
\begin{scope}
  \node (b) at (140:1.75) {};
  \path [postaction={decorate,decoration={raise=0ex,text along path, text align={center}, text={|\large|....}}}] (b)+(170:0.7cm) arc (170:210:0.8cm);
  \draw[Aactivity] (b)+(-150:0.2) -- ++(-150:0.8);
  \draw[Aactivity] (b)+(170:0.2) -- ++(170:0.8);
  \draw[Aactivity] (b)+(150:0.2) -- ++(150:0.8);
\end{scope}
\draw[Aactivity] (b)+(0:0.2) -- (0,0);
\draw[thick,fill=white] (b)+(0,0) circle (0.2cm);
\node at (160:2.5) {$m_1$};
\begin{scope}
  \path [postaction={decorate,decoration={raise=0ex,text along path, text align={center}, text={|\large|....}}}] (c)+(170:0.7cm) arc (170:210:0.8cm);
  \draw[Aactivity] (c)+(-150:0.2) -- ++(-150:0.8);
  \draw[Aactivity] (c)+(170:0.2) -- ++(170:0.8);
  \draw[Aactivity] (c)+(150:0.2) -- ++(150:0.8);
\end{scope}
\draw[Aactivity] (c)+(0:0.2) --  (0,0);
\draw[thick,fill=white] (c)+(0,0) circle (0.2cm);
\node at (207:2.6) {$m_\ell$};
\draw[density]  (s) -- (m);
}
\,\plaind t'
\elabel{cut_tree} \\
&= s \int_{t_0}^t \sin(\nu t')
\sum_{\ell=1}^n \ell g_\ell^{(0)}(t_0,t') \frac{1}{\ell!}
\times
\sum_{m_1=1, \ldots, m_\ell=1}^n
\binom{n}{m_1, \ldots, m_\ell} g_{m_1}^{(0)}(t',t) \cdots g_{m_\ell}^{(0)}(t',t)\, \plaind t' \nonumber \\
&= s \int_{0}^{t-t_0}\!\!\!\!\! \sin(\nu (t-t'))
\sum_{\ell=1}^n \ell g_\ell^{(0)}(0,t-t_0-t'') \frac{1}{\ell!}
\times\!\!\!\!\!\!
\sum_{m_1=1, \ldots, m_\ell=1}^n
\binom{n}{m_1, \ldots, m_\ell} g_{m_1}^{(0)}(0,t'') \cdots g_{m_\ell}^{(0)}(0,t'') \,\plaind t'' \, ,
\end{align}
\end{subequations}
\end{widetext}
where the last identity is obtained by the substitution $t''=t-t'$
and using the time-homogeneity
$g_\ell^{(0)}(t_0,t)=g_\ell^{(0)}(0,t-t_0)$. Defining the sum 
\begin{align}\elabel{def_S_sum}
&S(t-t_0,t'') =     
\sum_{\ell=1}^n \ell g_\ell^{(0)}(0,t-t_0-t'') \frac{1}{\ell!} \\
&\,\times\!\!\!\!
\sum_{m_1=1, \ldots, m_\ell=1}^n
\binom{n}{m_1, \ldots, m_\ell} g_{m_1}^{(0)}(0,t'') \cdots g_{m_\ell}^{(0)}(0,t'') \, ,
\nonumber
\end{align}
the first order perturbation can be written as
\begin{equation}
    g_n^{(1)}(t_0,t) = s \int_0^{t-t_0}
\!\! \sin(\nu (t-t'))
S(t-t_0,t'') \plaind t''\ .
\end{equation}
The sum 
$S(t-t_0,t'')$,
\Eref{def_S_sum}, 
would be equal to $g_n^{(0)}(t_0,t)$,
\Eref{gn0t2}, was it not for the factor $\ell$ 
in front of $g_\ell^{(0)}(0,t-t_0-t'')$, 
corresponding to the $\ell$ legs at time $t'$ to
choose from to insert the external perturbation. 
Nevertheless, the summation \Eref{def_S_sum} can be carried out
by using $g_n^{(0)}$ of \Eref{gn0} as a generating function. 
The term 
$\ell g_\ell^{(0)}(t_0,t)$ can be obtained 
from the time-derivative of the function $g_n^{(0)}(t_0,t)$ in \Eref{all_gn0},
\begin{equation}
\frac{\partial }{\partial t} g_\ell^{(0)}(t_0,t) = r g_\ell^{(0)}(t_0,t) \, \frac{\ell e^{-r(t-t_0)}-1}{1-e^{-r(t-t_0)}} \, ,
\end{equation}
which produces the identity
\begin{multline}
\elabel{lg}
\ell g_\ell^{(0)}(t_0,t) = \exp{r(t-t_0)}\\
\times\left(g_\ell^{(0)}(t_0,t) + \frac{1}{r}\left( 1-\exp{-r(t-t_0)}\right)\frac{\partial }{\partial t} g_\ell^{(0)}(t_0,t)  \right)
\, .
\end{multline}
Rewriting 
$\ell g_\ell^{(0)}(0,t-t_0-t'')=\ell g_\ell^{(0)}(t_0+t'',t)$
in \Eref{def_S_sum} in terms of
$\frac{\partial }{\partial t} g_\ell^{(0)}(t_0,t)$ and multiples of
$g_\ell^{(0)}(t_0,t)$ according to
\Eref{lg} 
allows the summation to be carried out, using \Eref{gn0t2} and \Eref{gn_time_derivative},
which eventually gives
\begin{multline}
S(t-t_0,t'') = g_n^{(0)}(0,t-t_0) \\
\times \left( 1+ (n-1)  \frac{e^{-rt''}-e^{-r(t-t_0)}}{1-e^{-r(t-t_0)}}\right)\, ,
\end{multline}
and therefore \Eref{moments-first-order-correction}.
For what follows,
this is best re-written as
\begin{multline}
\elabel{gn1_with_uv}
g_n^{(1)}(t_0,t)  
=  g_n^{(0)}(t_0,t) \Big(u(t_0,t)
+(n-1)v(t_0,t)\Big)
  \end{multline}
with
\begin{subequations}
\elabel{def_uv}
\begin{align}
u(t_0,t) &= s \int_0^{t-t_0}\!\! \sin(\nu (t-t')) \,\dint t'\nonumber\\
&= \frac{s}{\nu}(\cos(\nu t_0)-\cos(\nu t)) \\
v(t_0,t) &= s \int_0^{t-t_0} \!\!\!\sin(\nu (t-t')) \frac{e^{-rt'}-e^{-r(t-t_0)}}{1-e^{-r(t-t_0)}}\,\dint t' \nonumber\\
&= \frac{s}{1-e^{-r(t-t_0)}}
 \Big[ \frac{1}{\nu} \exp{-r(t-t_0)}(\cos(\nu t)-\cos(\nu t_0)) \nonumber\\
&\quad +\frac{1}{r^2+\nu^2} \Big( r\sin(\nu t) - \nu\cos(\nu t)\nonumber \\
&\quad\quad +\exp{-r(t-t_0)}(\nu\cos(\nu t_0) - r\sin(\nu t_0))\Big)\Big] \, .
\end{align}
\end{subequations}

\section{Further observables}
\subsection{Covariance}
\label{Sec:CovarianceDerivation}
\noindent The covariance is defined as\begin{align}
\text{Cov}(N(t_1),N(t_2))=\mathbb{E}[N(t_1)N(t_2)]-\mathbb{E}[N(t_1)]\mathbb{E}[N(t_2)],
\end{align}
where all expectations are conditioned to $N(0)=1$.
In the field theory, the term $\mathbb{E}[N(t_1)N(t_2)]$ is calculated as
\begin{align}
    &\mathbb{E}[N(t_1)N(t_2)]=\,\langle\phi(t_\text{max})\phi^\dagger(t_\text{min})\phi(t_\text{min})\phi^\dagger(0)\rangle \label{eq:cov-derivation} \\
    &\,=\langle\phi(t_\text{max})\phi(t_\text{min})\phitilde(0)\rangle+\langle\phi(t_\text{max})\widetilde\phi(t_\text{min})\phi(t_\text{min})\phitilde(0)\rangle,
    \nonumber
\end{align}
where $t_\text{max}=\text{max}\{t_1,t_2\}$ and $t_\text{min}=\text{min}\{t_1,t_2\}$. The first term of Eq.~\eqref{eq:cov-derivation} is a convolution similar to 
Eqs.~\eqref{eq:second-moment-derivation} and
\eqref{eq:secondmom_from_firstmom},
\begin{align}
    &\langle\phi(t_\text{max})\phi(t_\text{min})\phitilde(0)\rangle\hat=\,\tikz[baseline=-2.5pt]{\draw[very thick, color=red] (-0.7,0.5) node[above] {$\color{black}\scriptstyle t_\text{max}$} -- (0,0) -- (0.5,0.0) node[above] {$\color{black}\scriptstyle 0$};\draw[very thick, color=red](-0.5,-0.4) node[below] {$\color{black}\scriptstyle t_\text{min}$} -- (0,0);}\\
    &=2q_2\int\left\langle\phi(t_\text{max})\widetilde\phi(t')\right\rangle\left\langle\phi(t_\text{min})\widetilde\phi(t')\right\rangle\left\langle\phi(t')\widetilde\phi(0)\right\rangle\plaind t',
\end{align}
while the second term is a product of two first moments,
\begin{align}
    \langle\phi(t_\text{max})\widetilde\phi(t_\text{min})\phi(t_\text{min})\phitilde(0)\rangle\hat=&\,\tikz[baseline=-2.5pt]{\draw[very thick, color=red] (-0.9,0.0) node[above] {$\color{black}\scriptstyle t_\text{max}$} -- (-0.3,0);\draw[very thick, color=red] (-0.2,0.0) node[above] {$\color{black}\scriptstyle t_\text{min}$} -- (0.4,0) node[above] {$\color{black}\scriptstyle0$};} \\
    =\left\langle\phi(t_\text{max})\widetilde\phi(t_\text{min})\right\rangle&\left\langle\phi(t_\text{min})\widetilde\phi(0)\right\rangle
    \, .
\end{align}
At $t_1=t_2$, when $\ave{\phi(t_\text{max})\phitilde(t_\text{min})}=1$, \Eref{cov-derivation} 
recovers \Eref{2nd_moment}.

\subsection{Probability of survival to first order}
\label{Sec:SurvProb_derivation}
The survival probability $P_s(t_0,t)$ is determined via the probability $P(N(t)=0|N(t_0)=1)=1-P_s(t_0,t)$ 
that the system contains no particles at time $t$. Field-theoretically that is 
$\ave{\exp{-\phi(t)}\phi^\dagger(t_0)}$ 
\cite{Garcia-Millan2018,Pausch2019} so that
\begin{align}
P_s(t_0,t) 
& = 1-\ave{\exp{-\phi(t)}\phi^\dagger(t_0)} \\
& = - \sum_{n=1}^\infty \frac{(-)^n}{n!} \ave{\phi^n(t)\tildephi(t_0)} \\
& = \sum_{n=1}^\infty \frac{(-)^{n-1}}{n!} g_n(t_0,t) \,.
\end{align}
To zeroth order the summation is easily carried out, since $g^{(0)}_n(t_0,t)$, \Eref{gn0}, may be written as 
$g_n(t_0,t)=n!\alpha\beta^{n-1}$ with $\alpha=\exp{-r(t-t_0)}$ and $\beta=q_2(1-\exp{-r(t-t_0)})/r$,
\begin{equation}
\elabel{sum_Ps_in_alpha_beta}
P_s(t_0,t) = \alpha \sum_{m=0}^\infty (-\beta)^m + \OC(A) = \frac{\alpha}{1+\beta} + \OC(A)
\ ,
\end{equation}
as shown in Eq.~(42) in \cite{Garcia-Millan2018}.
Using the first-order corrected $g_n(t_0,t)$, \Eref{gn1_with_uv}, the summation does not change significantly as far as the term $A u(t_0,t)$ is concerned, but $A (n-1) v(t_0,t)$ requires some additional work. Recognising that \Eref{sum_Ps_in_alpha_beta} is effectively a generating function, the required sum is easily obtained from
\Eref{sum_Ps_in_alpha_beta},
\begin{multline}
P_s(t_0,t) 
= \frac{\alpha}{1+\beta} (1 + A u(t_0,t)) + \beta \frac{\partial}{\partial \beta} \frac{\alpha}{1+\beta} A v(t_0,t)
\end{multline}
To reduce clutter, we expand $u(t_0,t)$ and $v(t_0,t)$ only for the specific case of $t_0=0$. Using
\begin{equation}
P^{(0)}_s(0,t) =  \frac{\exp{-rt}}{1+\frac{q_2}{r}(1-\exp{-rt})}
\end{equation}
for the zeroth order approximation, the survival probability to first order reads
\begin{align}\elabel{P_s_full}
& P_s(0,t) 
=P^{(0)}_s(0,t)
 \Bigg\{
1+As
\Bigg[
\frac{1}{\nu}
+P^{(0)}_s(0,t)\frac{q_2 r}{(r^2+\nu^2)\nu} \nonumber\\
&- \cos(\nu t) \left(\frac{1}{\nu}+P^{(0)}_s(0,t)\frac{q_2}{r} \left(\frac{1}{\nu} - \frac{\nu}{\exp{-rt} (r^2+\nu^2)}\right) \right) \nonumber\\
&- \sin(\nu t) \left(P^{(0)}_s(0,t) \frac{q_2}{\exp{-rt} (r^2+\nu^2)} \right)
\Bigg]
\Bigg\} + \mO(A^2) \ .
\end{align}

\subsection{Avalanche shape to first order}
\label{Sec:ShapeDerivation}
In \cite{Garcia-Millan2018,Pausch2019}, it is shown that the avalanche shape (temporal profile)
of a branching process 
can be related to the incompletely normalised average
\begin{multline}
\Expected{N(t);N(T)=0|N(t_0)=1}
\\
=
\sum_N N P(N(t)=N,N(T)=0|N(t_0)=1)\ ,
\end{multline}
which is the average population size $N$ at time $t$
\emph{taken over the joint probability} 
$P(N(t)=N,N(T)=0|N(t_0)=1)$ 
of having size $N$ at time $t$ and size $0$ at time $T$, conditioned to having size $1$ at time $t_0$.
This quantity is easily captured in the field theory,
\begin{align}
\elabel{BP1eq47}
&\Expected{N(t);N(T)=0|N(t_0)=1}  \\
&=\ave{\exp{-\phi(T)}\phi^\dagger(t)\phi(t)\phi^\dagger(t_0)}\nonumber \\
&=
\ave{\phi(t)\phitilde(t_0)} +
\sum_{n=1}\frac{(-1)^n}{n!}\Bigg(
\ave{\phi^n(T)\phi(t)\tildephi(t_0)} \nonumber\\
&
\qquad\qquad+
\ave{\phi^n(T)\tildephi(t)\phi(t)\tildephi(t_0)}
\Bigg) \nonumber \, .
  \end{align}
Using identities (B2) and (B3) of 
\cite{Garcia-Millan2018} in the form
\begin{multline}
    \ave{\phi^n(T)\phi(t)\tildephi(t_0)}
    =
    \sum_{k=1}^n
    \sum_{m_1=1,\ldots,m_k=1}^n
    \binom{n}{m_1,\ldots, m_k}\\
\times
g_{m_1}(t,T)\ldots g_{m_k}(t,T)
g_{k+1}(t_0,t)\frac{1}{k!}
\end{multline}
and
\begin{multline}
    \!\!\!\!\!\!\!\ave{\phi^n(T)\tildephi(t)\phi(t)\tildephi(t_0)}
    =
    \sum_{k=1}^n
    \sum_{m_1=1,\ldots,m_k=1}^n
    \binom{n}{m_1,\ldots, m_k}\\
\times
g_{m_1}(t,T)\ldots g_{m_k}(t,T)
g_k(t_0,t)\frac{1}{(k-1)!} \, ,
\end{multline}
and
writing
$g_n(t_0,t) =n!\alpha\beta^{n-1}$ and
$g_n(t,T) =n!ab^{n-1}$,
\Eref{BP1eq47} can be written at $A=0$ 
after some tedious algebra
\begin{multline}
\mS:=\Expected{N(t);N(T)=0|N(t_0)=1}\elabel{shapeS0}
\\
= \alpha -\frac{a\alpha}{1+b\left(1+\frac{a\beta}{b}\right)}
\left[
1+\beta\left(2
- \frac{a(1+\beta)}{1+b\left(1+\frac{a\beta}{b}\right)}
\right)
\right]\,.
  \end{multline}
To zeroth order in $A$, we recover the result in \cite{Garcia-Millan2018} with
$\alpha=\exp{-r(t-t_0)}$, $\beta=\frac{q_2}{r}\left(1-\exp{-r(t-t_0)}\right)$,
as used in \Sref{SurvProb_derivation},
as well as
$a=\exp{-r(T-t)}$ and $b=\frac{q_2}{r}\left(1-\exp{-r(T-t)}\right)$.
To incorporate the first-order correction, we adjust $\alpha$, $\beta$, $a$ and $b$ so that $g_n(t_0,t)$ and 
$\ave{N(t),N(t_0)=1,N(T)=0}$ can be calculated order by order via generating functions.
 Defining
\begin{subequations}
\begin{align}
\alpha&=\exp{-r(t-t_0)}(1+A_1),\\
\beta&=\frac{q_2}{r}\left(1-\exp{-r(t-t_0)}\right)z_1,\\
a&=\exp{-r(T-t)}(1+A_2),\\
b&=\frac{q_2}{r}\left(1-\exp{-r(T-t)}\right)z_2,
  \end{align}
\end{subequations}
and the generating function $\mG=n!\alpha\beta^{n-1}$,
we recover $g_n(t_0,t)$ to first order according to \Eref{gn1_with_uv} 
in the form
\begin{align}
g_n(t_0,t)&=\Big\{
\mG+A\Big(u(t_0,t)\frac{\partial \mG}{\partial A_1} \\
&+v(t_0,t)\frac{\partial \mG}{\partial z_1} \Big)
\Big\} \Big|_{A_1=0, z_1=1}  + \mO\left(A^2\right) \, , \nonumber
  \end{align}
and using the generating function
$\mS$ 
as defined in \Eref{shapeS0},
similarly for the avalanche shape,
\begin{align}
\elabel{ava_shape_0}
&\Expected{N(t);N(T)=0|N(t_0)=1}\\
&=  \left\{\mS
+ A\left(
u(t_0,t)\frac{\partial \mS}{\partial A_1} +u(t,T)\frac{\partial \mS}{\partial A_2}
+v(t_0,t)\frac{\partial \mS}{\partial z_1}  \right.\right. \nonumber\\
&\quad\left.\left.
+v(t,T)\frac{\partial \mS}{\partial z_2} 
\right)
\right\} \Big|_{A_1=0, z_1=1, A_2=0, z_2=1} + \mO\left(A^2\right) \, .
\nonumber
\end{align}
At this point, it is worth noting that \Eref{ava_shape_0} is a general expression whose dependence on the time-dependent extinction rate lies in the functions $u$ and $v$. Hence, this expression of the avalanche shape holds for different time-dependent extinction rates as long as
$u$ and $v$ are calculated along the lines
of \Eref{def_uv}
As indicated earlier, the expression in \Eref{ava_shape_0} 
is the expectation of $N$ over the 
\emph{joint} probability of $N(t)=N$
and $N(T)=0$. This includes all trajectories 
that become extinct before time $T$.
To calculate the avalanche shape of those instances that become extinct exactly at time $T$, we need the expectation
of $N(t)$ \emph{conditioned} to $N(T)=0$ and $N(t)>0$ for 
$t<T$,
\begin{equation}
    V(t,T)= \Expected{N(t)|N(T>t>t_0)>0,N(T)=0,N(t_0)=1} \,.
\end{equation}
Assuming $\plaind T>0$, those trajectories that have $N(T+\plaind T)=0$
but $N(T)\neq0$, have the desired feature of going extinct
precisely at time $T$.
Since $N(T)=0$ implies $N(T+\plaind T)=0$,
the shape $V(t,T)$  is equal to the expectation
$\Expected{N(t);N(T)=0|N(t_0)=1}$  differentiated with respect
to $T$, and normalised by the probability density of
extinction at exactly time $T$, 
\begin{equation}
    V(t,T)=\frac{
    \frac{\plaind}{\plaind T} \Expected{N(t);N(T)=0|N(t_0)=1}}{ - \frac{\plaind}{\plaind T} P_s(0,T)} \, ,
\end{equation}
which can be calculated to first order in $A$ by using
\Erefs{P_s_full} and \eref{ava_shape_0}.
However, numerically it is much easier to normalise the avalanche 
shape $V(t,T)$ by the area under its curve
\begin{equation}\elabel{def_V_1}
    V_1(t,T) = \frac{V(t,T)}{\int_0^1 V(\tau T,T) \plaind \tau} \ ,
\end{equation}
which is what is shown in the plots in Figs.~\ref{fig:shape01} and \ref{fig:largeA_shape}.
\bibliography{books,articles}
\end{document}

%% file: mydiagrams.tex
\usepackage{tikz}
\usetikzlibrary{decorations.pathmorphing}
\usetikzlibrary{decorations.markings}
\usetikzlibrary{arrows,shapes,snakes,automata,backgrounds,petri}
\usetikzlibrary{calc}
\usetikzlibrary{patterns}
\usetikzlibrary{decorations.text}
\tikzset{
xxtsubstrate/.style={decorate, 
line width=1pt,
draw=olive, 
decoration=snake, 
segment amplitude=0.75mm, 
line after snake=0.25mm,
line before snake=0.25mm
},
tsubstrate/.style={decorate, 
line width=1pt,
draw=olive, 
decoration=snake, 
segment amplitude=0.5mm, 
segment length=5pt,
segment amplitude=0.2mm, 
line after snake=1mm,
line before snake=1mm
},
Bsubstrate/.style={decorate, 
line width=1pt,
draw=olive, 
decoration=snake,
segment length=5pt,
segment aspect=0,
segment amplitude=0.5mm, 
line after snake=0mm,
line before snake=0mm
},
substrate/.style={decorate, 
line width=1pt,
draw=olive, 
decoration=snake, 
segment length=5pt,
segment amplitude=0.5mm, 
line after snake=0.5mm,
line before snake=0.5mm
},
activity/.style={very thick,draw=red,postaction={decorate},
decoration={markings,mark=at position .5 with
{\arrow[draw=red]{>}}}},
tactivity/.style={thick,draw=red,postaction={decorate},
decoration={markings,mark=at position .5 with
{\arrow[draw=red]{>}}}},
tEPSactivity/.style={thick,draw=red,postaction={decorate},
decoration={markings,mark=at position .55 with
{\arrow[draw=red]{>}}}},
tAactivity/.style={thick,draw=red},
Aactivity/.style={very thick,draw=red},
Cactivity/.style={very thick,draw=blue,decorate,decoration={snake}},
Bactivity/.style={very thick,draw=blue,dashed},
tSactivity/.style={thick,draw=red,postaction={decorate},
decoration={markings,mark=at position .7 with
{\arrow[draw=red]{>}}}},
Sactivity/.style={very thick,draw=red,postaction={decorate},
decoration={markings,mark=at position .7 with
{\arrow[draw=red]{>}}}},
density/.style={ 
line width=1.5pt,
draw=black,
densely dashed,
segment length=5pt,
segment amplitude=0.5mm, 
}
}

%% file: preamble.tex
\usepackage[colorlinks=true,linkcolor=blue, urlcolor = black, citecolor = blue, anchorcolor = blue]{hyperref}

\usepackage{graphicx}
\usepackage[tight]{subfigure}
\subfiguretopcaptrue
\usepackage{multirow}
\usepackage{subeqnarray}
\usepackage{wasysym}
\usepackage{eufrak}

\newcommand{\ave}[1]{\left\langle #1 \right\rangle}

\newcommand{\plaind}{\mathrm{d}}

\newcommand{\tildephi}{\widetilde\phi}

\newcommand{\Eref}[1]{Eq.~(\ref{eq:#1})}
\newcommand{\Esref}[1]{Eqs.~(\ref{eq:#1})}
\newcommand{\eref}[1]{(\ref{eq:#1})}

\newcommand{\Sref}[1]{Section~\ref{sec:#1}}

\newcommand{\elabel}[1]{\label{eq:#1}}

\newcommand{\dint}[1]{\mathchoice{\!\plaind#1\,}{\!\plaind#1\,}{\!\plaind#1\,}{\!\plaind#1\,}}

\newcommand{\Exp}[1]{\operatorname{exp}\!\left(#1\right)}
\renewcommand{\exp}[1]{\mathchoice{e^{#1}}{\operatorname{exp}\!\left(#1\right)}{\operatorname{exp}\!\left(#1\right)}{\operatorname{exp}\!\left(#1\right)}}
\newcommand{\corresponding}{\hat{=}}

\newcommand{\mG}{\mathcal{G}}
\newcommand{\mO}{\mathcal{O}}

\usepackage{xspace}
\newcommand{\latin}[1]{{\it #1}}
\newcommand{\ie}{\latin{i.e.}\@\xspace}
\newcommand{\eg}{\latin{e.g.}\@\xspace}
\newcommand{\cf}{\latin{cf.}\@\xspace}

\newcommand{\etal}{\latin{et~al}.\@\xspace}

\newcommand{\Erefs}[1]{Eqs.~(\ref{eq:#1})}

%% file: main.bbl
\begin{thebibliography}{50}%
\makeatletter
\providecommand \@ifxundefined [1]{%
 \@ifx{#1\undefined}
}%
\providecommand \@ifnum [1]{%
 \ifnum #1\expandafter \@firstoftwo
 \else \expandafter \@secondoftwo
 \fi
}%
\providecommand \@ifx [1]{%
 \ifx #1\expandafter \@firstoftwo
 \else \expandafter \@secondoftwo
 \fi
}%
\providecommand \natexlab [1]{#1}%
\providecommand \enquote  [1]{``#1''}%
\providecommand \bibnamefont  [1]{#1}%
\providecommand \bibfnamefont [1]{#1}%
\providecommand \citenamefont [1]{#1}%
\providecommand \href@noop [0]{\@secondoftwo}%
\providecommand \href [0]{\begingroup \@sanitize@url \@href}%
\providecommand \@href[1]{\@@startlink{#1}\@@href}%
\providecommand \@@href[1]{\endgroup#1\@@endlink}%
\providecommand \@sanitize@url [0]{\catcode `\\12\catcode `\$12\catcode
  `\&12\catcode `\#12\catcode `\^12\catcode `\_12\catcode `\%12\relax}%
\providecommand \@@startlink[1]{}%
\providecommand \@@endlink[0]{}%
\providecommand \url  [0]{\begingroup\@sanitize@url \@url }%
\providecommand \@url [1]{\endgroup\@href {#1}{\urlprefix }}%
\providecommand \urlprefix  [0]{URL }%
\providecommand \Eprint [0]{\href }%
\providecommand \doibase [0]{http://dx.doi.org/}%
\providecommand \selectlanguage [0]{\@gobble}%
\providecommand \bibinfo  [0]{\@secondoftwo}%
\providecommand \bibfield  [0]{\@secondoftwo}%
\providecommand \translation [1]{[#1]}%
\providecommand \BibitemOpen [0]{}%
\providecommand \bibitemStop [0]{}%
\providecommand \bibitemNoStop [0]{.\EOS\space}%
\providecommand \EOS [0]{\spacefactor3000\relax}%
\providecommand \BibitemShut  [1]{\csname bibitem#1\endcsname}%
\let\auto@bib@innerbib\@empty
\bibitem [{\citenamefont {Dayan}\ and\ \citenamefont
  {Abbott}(2001)}]{Dayan2001}%
  \BibitemOpen
  \bibfield  {author} {\bibinfo {author} {\bibfnamefont {P.}~\bibnamefont
  {Dayan}}\ and\ \bibinfo {author} {\bibfnamefont {L.}~\bibnamefont {Abbott}},\
  }\href@noop {} {\emph {\bibinfo {title} {Theoretical Neuroscience}}}\
  (\bibinfo  {publisher} {The MIT Press},\ \bibinfo {address} {Cambridge, MA,
  USA},\ \bibinfo {year} {2001})\BibitemShut {NoStop}%
\bibitem [{\citenamefont {Plenz}\ and\ \citenamefont
  {Aertsen}(1996)}]{Plenz1996}%
  \BibitemOpen
  \bibfield  {author} {\bibinfo {author} {\bibfnamefont {D.}~\bibnamefont
  {Plenz}}\ and\ \bibinfo {author} {\bibfnamefont {A.}~\bibnamefont
  {Aertsen}},\ }\href {\doibase https://doi.org/10.1016/0306-4522(95)00406-8}
  {\bibfield  {journal} {\bibinfo  {journal} {Neuroscience}\ }\textbf {\bibinfo
  {volume} {70}},\ \bibinfo {pages} {861 } (\bibinfo {year}
  {1996})}\BibitemShut {NoStop}%
\bibitem [{\citenamefont {Plenz}\ and\ \citenamefont
  {Kitai}(1998)}]{Plenz1998}%
  \BibitemOpen
  \bibfield  {author} {\bibinfo {author} {\bibfnamefont {D.}~\bibnamefont
  {Plenz}}\ and\ \bibinfo {author} {\bibfnamefont {S.~T.}\ \bibnamefont
  {Kitai}},\ }\href {\doibase 10.1523/JNEUROSCI.18-01-00266.1998} {\bibfield
  {journal} {\bibinfo  {journal} {J Neuroscience}\ }\textbf {\bibinfo {volume}
  {18}},\ \bibinfo {pages} {266} (\bibinfo {year} {1998})}\BibitemShut
  {NoStop}%
\bibitem [{\citenamefont {Karpiak}\ and\ \citenamefont
  {Plenz}(2002)}]{Karpiak2002}%
  \BibitemOpen
  \bibfield  {author} {\bibinfo {author} {\bibfnamefont {V.~C.}\ \bibnamefont
  {Karpiak}}\ and\ \bibinfo {author} {\bibfnamefont {D.}~\bibnamefont
  {Plenz}},\ }\href {\doibase 10.1002/0471142301.ns0615s19} {\bibfield
  {journal} {\bibinfo  {journal} {Curr. Protoc. Neurosci.}\ }\textbf {\bibinfo
  {volume} {19}},\ \bibinfo {pages} {6.15.1} (\bibinfo {year}
  {2002})}\BibitemShut {NoStop}%
\bibitem [{\citenamefont {Beggs}\ and\ \citenamefont
  {Plenz}(2003)}]{BeggsPlenz:2003}%
  \BibitemOpen
  \bibfield  {author} {\bibinfo {author} {\bibfnamefont {J.~M.}\ \bibnamefont
  {Beggs}}\ and\ \bibinfo {author} {\bibfnamefont {D.}~\bibnamefont {Plenz}},\
  }\href@noop {} {\bibfield  {journal} {\bibinfo  {journal} {J. Neurosci.}\
  }\textbf {\bibinfo {volume} {23}},\ \bibinfo {pages} {11167} (\bibinfo {year}
  {2003})}\BibitemShut {NoStop}%
\bibitem [{\citenamefont {Beggs}\ and\ \citenamefont
  {Plenz}(2004)}]{BeggsPlenz:2004}%
  \BibitemOpen
  \bibfield  {author} {\bibinfo {author} {\bibfnamefont {J.~M.}\ \bibnamefont
  {Beggs}}\ and\ \bibinfo {author} {\bibfnamefont {D.}~\bibnamefont {Plenz}},\
  }\href@noop {} {\bibfield  {journal} {\bibinfo  {journal} {J. Neurosci.}\
  }\textbf {\bibinfo {volume} {24}},\ \bibinfo {pages} {5216} (\bibinfo {year}
  {2004})}\BibitemShut {NoStop}%
\bibitem [{\citenamefont {Priesemann}\ \emph {et~al.}(2009)\citenamefont
  {Priesemann}, \citenamefont {Munk},\ and\ \citenamefont
  {Wibral}}]{Priesemann:2009}%
  \BibitemOpen
  \bibfield  {author} {\bibinfo {author} {\bibfnamefont {V.}~\bibnamefont
  {Priesemann}}, \bibinfo {author} {\bibfnamefont {M.~H.}\ \bibnamefont
  {Munk}}, \ and\ \bibinfo {author} {\bibfnamefont {M.}~\bibnamefont
  {Wibral}},\ }\href {\doibase 10.1186/1471-2202-10-40} {\bibfield  {journal}
  {\bibinfo  {journal} {BMC Neurosci.}\ }\textbf {\bibinfo {volume} {10}},\
  \bibinfo {pages} {1} (\bibinfo {year} {2009})}\BibitemShut {NoStop}%
\bibitem [{\citenamefont {Priesemann}\ \emph {et~al.}(2013)\citenamefont
  {Priesemann}, \citenamefont {Valderrama}, \citenamefont {Wibral},\ and\
  \citenamefont {Quyen}}]{Priesemann2013}%
  \BibitemOpen
  \bibfield  {author} {\bibinfo {author} {\bibfnamefont {V.}~\bibnamefont
  {Priesemann}}, \bibinfo {author} {\bibfnamefont {M.}~\bibnamefont
  {Valderrama}}, \bibinfo {author} {\bibfnamefont {M.}~\bibnamefont {Wibral}},
  \ and\ \bibinfo {author} {\bibfnamefont {M.~L.~V.}\ \bibnamefont {Quyen}},\
  }\href {\doibase 10.1371/journal.pcbi.1002985} {\bibfield  {journal}
  {\bibinfo  {journal} {PLoS Comput. Biol.}\ }\textbf {\bibinfo {volume} {9}}
  (\bibinfo {year} {2013}),\ 10.1371/journal.pcbi.1002985}\BibitemShut
  {NoStop}%
\bibitem [{\citenamefont {Wagenaar}\ \emph {et~al.}(2006)\citenamefont
  {Wagenaar}, \citenamefont {Pine},\ and\ \citenamefont
  {Potter}}]{Wagenaar2006}%
  \BibitemOpen
  \bibfield  {author} {\bibinfo {author} {\bibfnamefont {D.~A.}\ \bibnamefont
  {Wagenaar}}, \bibinfo {author} {\bibfnamefont {J.}~\bibnamefont {Pine}}, \
  and\ \bibinfo {author} {\bibfnamefont {S.~M.}\ \bibnamefont {Potter}},\
  }\href {\doibase 10.1186/1471-2202-7-11} {\bibfield  {journal} {\bibinfo
  {journal} {BMC Neuroscience}\ }\textbf {\bibinfo {volume} {7}},\ \bibinfo
  {pages} {11} (\bibinfo {year} {2006})}\BibitemShut {NoStop}%
\bibitem [{\citenamefont {Beggs}(2008)}]{Beggs:2008}%
  \BibitemOpen
  \bibfield  {author} {\bibinfo {author} {\bibfnamefont {J.~M.}\ \bibnamefont
  {Beggs}},\ }\href@noop {} {\bibfield  {journal} {\bibinfo  {journal} {Phil.
  Trans. R. Soc. A}\ }\textbf {\bibinfo {volume} {366}},\ \bibinfo {pages}
  {329} (\bibinfo {year} {2008})}\BibitemShut {NoStop}%
\bibitem [{\citenamefont {Brochini}\ \emph {et~al.}(2016)\citenamefont
  {Brochini}, \citenamefont {de~Andrade~Costa}, \citenamefont {Abadi},
  \citenamefont {Roque}, \citenamefont {Stolfi},\ and\ \citenamefont
  {Kinouchi}}]{Brochini2016}%
  \BibitemOpen
  \bibfield  {author} {\bibinfo {author} {\bibfnamefont {L.}~\bibnamefont
  {Brochini}}, \bibinfo {author} {\bibfnamefont {A.}~\bibnamefont
  {de~Andrade~Costa}}, \bibinfo {author} {\bibfnamefont {M.}~\bibnamefont
  {Abadi}}, \bibinfo {author} {\bibfnamefont {A.~C.}\ \bibnamefont {Roque}},
  \bibinfo {author} {\bibfnamefont {J.}~\bibnamefont {Stolfi}}, \ and\ \bibinfo
  {author} {\bibfnamefont {O.}~\bibnamefont {Kinouchi}},\ }\href {\doibase
  10.1038/srep35831} {\bibfield  {journal} {\bibinfo  {journal} {Sci. Rep.}\
  }\textbf {\bibinfo {volume} {6}} (\bibinfo {year} {2016}),\
  10.1038/srep35831}\BibitemShut {NoStop}%
\bibitem [{\citenamefont {Haldeman}\ and\ \citenamefont
  {Beggs}(2005)}]{Haldeman2005}%
  \BibitemOpen
  \bibfield  {author} {\bibinfo {author} {\bibfnamefont {C.}~\bibnamefont
  {Haldeman}}\ and\ \bibinfo {author} {\bibfnamefont {J.~M.}\ \bibnamefont
  {Beggs}},\ }\href {\doibase 10.1103/PhysRevLett.94.058101} {\bibfield
  {journal} {\bibinfo  {journal} {Phys. Rev. Lett.}\ }\textbf {\bibinfo
  {volume} {94}},\ \bibinfo {pages} {058101} (\bibinfo {year}
  {2005})}\BibitemShut {NoStop}%
\bibitem [{\citenamefont {Williams-Garc\'{\i}a}\ \emph
  {et~al.}(2014)\citenamefont {Williams-Garc\'{\i}a}, \citenamefont {Moore},
  \citenamefont {Beggs},\ and\ \citenamefont {Ortiz}}]{Williams-Garcia2014}%
  \BibitemOpen
  \bibfield  {author} {\bibinfo {author} {\bibfnamefont {R.~V.}\ \bibnamefont
  {Williams-Garc\'{\i}a}}, \bibinfo {author} {\bibfnamefont {M.}~\bibnamefont
  {Moore}}, \bibinfo {author} {\bibfnamefont {J.~M.}\ \bibnamefont {Beggs}}, \
  and\ \bibinfo {author} {\bibfnamefont {G.}~\bibnamefont {Ortiz}},\ }\href
  {\doibase 10.1103/PhysRevE.90.062714} {\bibfield  {journal} {\bibinfo
  {journal} {Phys. Rev. E}\ }\textbf {\bibinfo {volume} {90}},\ \bibinfo
  {pages} {062714} (\bibinfo {year} {2014})}\BibitemShut {NoStop}%
\bibitem [{\citenamefont {Wilting}\ and\ \citenamefont
  {Priesemann}(2019{\natexlab{a}})}]{Wilting2019}%
  \BibitemOpen
  \bibfield  {author} {\bibinfo {author} {\bibfnamefont {J.}~\bibnamefont
  {Wilting}}\ and\ \bibinfo {author} {\bibfnamefont {V.}~\bibnamefont
  {Priesemann}},\ }\href {\doibase 10.1093/cercor/bhz049} {\bibfield  {journal}
  {\bibinfo  {journal} {Cereb. Cortex}\ }\textbf {\bibinfo {volume} {29}},\
  \bibinfo {pages} {2759} (\bibinfo {year} {2019}{\natexlab{a}})}\BibitemShut
  {NoStop}%
\bibitem [{\citenamefont {Pruessner}(2012)}]{Pruessner:2012:Book}%
  \BibitemOpen
  \bibfield  {author} {\bibinfo {author} {\bibfnamefont {G.}~\bibnamefont
  {Pruessner}},\ }\href@noop {} {\emph {\bibinfo {title} {Self-Organised
  Criticality}}}\ (\bibinfo  {publisher} {Cambridge University Press},\
  \bibinfo {address} {Cambridge, UK},\ \bibinfo {year} {2012})\BibitemShut
  {NoStop}%
\bibitem [{\citenamefont {Garcia-Millan}\ \emph {et~al.}(2018)\citenamefont
  {Garcia-Millan}, \citenamefont {Pausch}, \citenamefont {Walter},\ and\
  \citenamefont {Pruessner}}]{Garcia-Millan2018}%
  \BibitemOpen
  \bibfield  {author} {\bibinfo {author} {\bibfnamefont {R.}~\bibnamefont
  {Garcia-Millan}}, \bibinfo {author} {\bibfnamefont {J.}~\bibnamefont
  {Pausch}}, \bibinfo {author} {\bibfnamefont {B.}~\bibnamefont {Walter}}, \
  and\ \bibinfo {author} {\bibfnamefont {G.}~\bibnamefont {Pruessner}},\ }\href
  {\doibase DOI: 10.1103/PhysRevE.98.062107} {\bibfield  {journal} {\bibinfo
  {journal} {Phys. Rev. E}\ }\textbf {\bibinfo {volume} {98}},\ \bibinfo
  {pages} {062107} (\bibinfo {year} {2018})}\BibitemShut {NoStop}%
\bibitem [{\citenamefont {Watson}\ and\ \citenamefont
  {Galton}(1875)}]{Watson:1875}%
  \BibitemOpen
  \bibfield  {author} {\bibinfo {author} {\bibfnamefont {H.}~\bibnamefont
  {Watson}}\ and\ \bibinfo {author} {\bibfnamefont {F.}~\bibnamefont
  {Galton}},\ }\href@noop {} {\bibfield  {journal} {\bibinfo  {journal} {Royal
  Anthropol. Inst. G. B. Irel.}\ }\textbf {\bibinfo {volume} {4}},\ \bibinfo
  {pages} {138} (\bibinfo {year} {1875})}\BibitemShut {NoStop}%
\bibitem [{\citenamefont {Harris}(1963)}]{Harris:1963}%
  \BibitemOpen
  \bibfield  {author} {\bibinfo {author} {\bibfnamefont {T.~E.}\ \bibnamefont
  {Harris}},\ }\href@noop {} {\emph {\bibinfo {title} {The Theory of Branching
  Processes}}}\ (\bibinfo  {publisher} {Springer-Verlag},\ \bibinfo {address}
  {Berlin, Germany},\ \bibinfo {year} {1963})\BibitemShut {NoStop}%
\bibitem [{\citenamefont {Athreya}\ and\ \citenamefont
  {Ney}(1972)}]{AthreyaNey:1972}%
  \BibitemOpen
  \bibfield  {author} {\bibinfo {author} {\bibfnamefont {K.~B.}\ \bibnamefont
  {Athreya}}\ and\ \bibinfo {author} {\bibfnamefont {P.~E.}\ \bibnamefont
  {Ney}},\ }\href@noop {} {\emph {\bibinfo {title} {Branching processes}}},\
  \bibinfo {series} {Grundlehren der mathematischen Wissenschaften}, Vol.\
  \bibinfo {volume} {196}\ (\bibinfo  {publisher} {Springer-Verlag},\ \bibinfo
  {address} {Berlin, Germany},\ \bibinfo {year} {1972})\BibitemShut {NoStop}%
\bibitem [{\citenamefont {P\'azsit}\ and\ \citenamefont
  {P\'al}(2007)}]{Pazsit:2007}%
  \BibitemOpen
  \bibfield  {author} {\bibinfo {author} {\bibfnamefont {I.}~\bibnamefont
  {P\'azsit}}\ and\ \bibinfo {author} {\bibfnamefont {L.}~\bibnamefont
  {P\'al}},\ }\href@noop {} {\emph {\bibinfo {title} {Neutron Fluctuations: A
  Treatise on the Physics of Branching Processes}}}\ (\bibinfo  {publisher}
  {Elsevier},\ \bibinfo {address} {Amsterdam},\ \bibinfo {year}
  {2007})\BibitemShut {NoStop}%
\bibitem [{\citenamefont {Williams}(2013)}]{Williams:2013}%
  \BibitemOpen
  \bibfield  {author} {\bibinfo {author} {\bibfnamefont {M.}~\bibnamefont
  {Williams}},\ }\href@noop {} {\emph {\bibinfo {title} {Random Processes in
  nuclear reactors}}}\ (\bibinfo  {publisher} {Elsevier},\ \bibinfo {address}
  {Amsterdam},\ \bibinfo {year} {2013})\BibitemShut {NoStop}%
\bibitem [{\citenamefont {Marzocchi}\ and\ \citenamefont
  {Lombardi}(2008)}]{Marzocchi:2008}%
  \BibitemOpen
  \bibfield  {author} {\bibinfo {author} {\bibfnamefont {W.}~\bibnamefont
  {Marzocchi}}\ and\ \bibinfo {author} {\bibfnamefont {A.}~\bibnamefont
  {Lombardi}},\ }\href {\doibase 10.1029/2007JB005472} {\bibfield  {journal}
  {\bibinfo  {journal} {J. Geophys. Res.}\ }\textbf {\bibinfo {volume} {113}}
  (\bibinfo {year} {2008}),\ 10.1029/2007JB005472}\BibitemShut {NoStop}%
\bibitem [{\citenamefont {Lee}\ \emph {et~al.}(2004)\citenamefont {Lee},
  \citenamefont {Goh}, \citenamefont {Kahng},\ and\ \citenamefont
  {Kim}}]{Lee:2004}%
  \BibitemOpen
  \bibfield  {author} {\bibinfo {author} {\bibfnamefont {D.}~\bibnamefont
  {Lee}}, \bibinfo {author} {\bibfnamefont {K.-I.}\ \bibnamefont {Goh}},
  \bibinfo {author} {\bibfnamefont {B.}~\bibnamefont {Kahng}}, \ and\ \bibinfo
  {author} {\bibfnamefont {D.}~\bibnamefont {Kim}},\ }\href@noop {} {\bibfield
  {journal} {\bibinfo  {journal} {J. Korean Phys. Soc.}\ }\textbf {\bibinfo
  {volume} {44}},\ \bibinfo {pages} {633} (\bibinfo {year} {2004})}\BibitemShut
  {NoStop}%
\bibitem [{\citenamefont {Simkin}\ and\ \citenamefont
  {Roychowdhury}(2010)}]{Simkin2010}%
  \BibitemOpen
  \bibfield  {author} {\bibinfo {author} {\bibfnamefont {M.}~\bibnamefont
  {Simkin}}\ and\ \bibinfo {author} {\bibfnamefont {V.}~\bibnamefont
  {Roychowdhury}},\ }\href {\doibase 10.1016/j.physrep.2010.12.004} {\bibfield
  {journal} {\bibinfo  {journal} {Phys. Rep.}\ }\textbf {\bibinfo {volume}
  {502}},\ \bibinfo {pages} {1} (\bibinfo {year} {2010})}\BibitemShut {NoStop}%
\bibitem [{\citenamefont {Durrett}(2015)}]{durrett2015branching}%
  \BibitemOpen
  \bibfield  {author} {\bibinfo {author} {\bibfnamefont {R.}~\bibnamefont
  {Durrett}},\ }in\ \href@noop {} {\emph {\bibinfo {booktitle} {Branching
  Process Models of Cancer}}}\ (\bibinfo  {publisher} {Springer},\ \bibinfo
  {year} {2015})\ pp.\ \bibinfo {pages} {1--63}\BibitemShut {NoStop}%
\bibitem [{\citenamefont {Gleeson}\ and\ \citenamefont
  {Durrett}(2017)}]{GleesonDurrett:2017}%
  \BibitemOpen
  \bibfield  {author} {\bibinfo {author} {\bibfnamefont {J.~P.}\ \bibnamefont
  {Gleeson}}\ and\ \bibinfo {author} {\bibfnamefont {R.}~\bibnamefont
  {Durrett}},\ }\href@noop {} {\bibfield  {journal} {\bibinfo  {journal} {Nat.
  Com.}\ }\textbf {\bibinfo {volume} {8}},\ \bibinfo {pages} {1227} (\bibinfo
  {year} {2017})}\BibitemShut {NoStop}%
\bibitem [{\citenamefont {Seshadri}\ \emph {et~al.}(2018)\citenamefont
  {Seshadri}, \citenamefont {Klaus}, \citenamefont {Winkowski}, \citenamefont
  {Kanold},\ and\ \citenamefont {Plenz}}]{seshadri2018altered}%
  \BibitemOpen
  \bibfield  {author} {\bibinfo {author} {\bibfnamefont {S.}~\bibnamefont
  {Seshadri}}, \bibinfo {author} {\bibfnamefont {A.}~\bibnamefont {Klaus}},
  \bibinfo {author} {\bibfnamefont {D.~E.}\ \bibnamefont {Winkowski}}, \bibinfo
  {author} {\bibfnamefont {P.~O.}\ \bibnamefont {Kanold}}, \ and\ \bibinfo
  {author} {\bibfnamefont {D.}~\bibnamefont {Plenz}},\ }\href@noop {}
  {\bibfield  {journal} {\bibinfo  {journal} {Translational psychiatry}\
  }\textbf {\bibinfo {volume} {8}},\ \bibinfo {pages} {3} (\bibinfo {year}
  {2018})}\BibitemShut {NoStop}%
\bibitem [{\citenamefont {Poil}\ \emph {et~al.}(2008)\citenamefont {Poil},
  \citenamefont {can Ooyen},\ and\ \citenamefont
  {Linkenkaer-Hansen}}]{Poil2008}%
  \BibitemOpen
  \bibfield  {author} {\bibinfo {author} {\bibfnamefont {S.-S.}\ \bibnamefont
  {Poil}}, \bibinfo {author} {\bibfnamefont {A.}~\bibnamefont {can Ooyen}}, \
  and\ \bibinfo {author} {\bibfnamefont {K.}~\bibnamefont
  {Linkenkaer-Hansen}},\ }\href {\doibase 10.1002/hbm.20590} {\bibfield
  {journal} {\bibinfo  {journal} {Hum. brain mapp.}\ }\textbf {\bibinfo
  {volume} {29}},\ \bibinfo {pages} {770} (\bibinfo {year} {2008})}\BibitemShut
  {NoStop}%
\bibitem [{\citenamefont {Wilting}\ \emph {et~al.}(2018)\citenamefont
  {Wilting}, \citenamefont {Dehning}, \citenamefont {Neto}, \citenamefont
  {Rudelt}, \citenamefont {Wibral}, \citenamefont {Zierenberg},\ and\
  \citenamefont {Priesemann}}]{Wilting2018}%
  \BibitemOpen
  \bibfield  {author} {\bibinfo {author} {\bibfnamefont {J.}~\bibnamefont
  {Wilting}}, \bibinfo {author} {\bibfnamefont {J.}~\bibnamefont {Dehning}},
  \bibinfo {author} {\bibfnamefont {J.~P.}\ \bibnamefont {Neto}}, \bibinfo
  {author} {\bibfnamefont {L.}~\bibnamefont {Rudelt}}, \bibinfo {author}
  {\bibfnamefont {M.}~\bibnamefont {Wibral}}, \bibinfo {author} {\bibfnamefont
  {J.}~\bibnamefont {Zierenberg}}, \ and\ \bibinfo {author} {\bibfnamefont
  {V.}~\bibnamefont {Priesemann}},\ }\href {\doibase 10.3389/fnsys.2018.00055}
  {\bibfield  {journal} {\bibinfo  {journal} {Front. Syst. Neurosci.}\ }\textbf
  {\bibinfo {volume} {12}} (\bibinfo {year} {2018}),\
  10.3389/fnsys.2018.00055}\BibitemShut {NoStop}%
\bibitem [{\citenamefont {Wilting}\ and\ \citenamefont
  {Priesemann}(2018)}]{Wilting2018b}%
  \BibitemOpen
  \bibfield  {author} {\bibinfo {author} {\bibfnamefont {J.}~\bibnamefont
  {Wilting}}\ and\ \bibinfo {author} {\bibfnamefont {V.}~\bibnamefont
  {Priesemann}},\ }\href {\doibase 10.1038/s41467-018-04725-4} {\bibfield
  {journal} {\bibinfo  {journal} {Nat. Commun.}\ }\textbf {\bibinfo {volume}
  {9}},\ \bibinfo {pages} {2325} (\bibinfo {year} {2018})}\BibitemShut
  {NoStop}%
\bibitem [{\citenamefont {Timme}\ \emph {et~al.}(2016)\citenamefont {Timme},
  \citenamefont {Marshall}, \citenamefont {Bennett}, \citenamefont {Ripp},
  \citenamefont {Lautzenhiser},\ and\ \citenamefont {Beggs}}]{Timme2016}%
  \BibitemOpen
  \bibfield  {author} {\bibinfo {author} {\bibfnamefont {N.~M.}\ \bibnamefont
  {Timme}}, \bibinfo {author} {\bibfnamefont {N.~J.}\ \bibnamefont {Marshall}},
  \bibinfo {author} {\bibfnamefont {N.}~\bibnamefont {Bennett}}, \bibinfo
  {author} {\bibfnamefont {M.}~\bibnamefont {Ripp}}, \bibinfo {author}
  {\bibfnamefont {E.}~\bibnamefont {Lautzenhiser}}, \ and\ \bibinfo {author}
  {\bibfnamefont {J.~M.}\ \bibnamefont {Beggs}},\ }\href {\doibase
  10.3389/fphys.2016.00425} {\bibfield  {journal} {\bibinfo  {journal} {Front.
  Physiol.}\ }\textbf {\bibinfo {volume} {7}},\ \bibinfo {pages} {425}
  (\bibinfo {year} {2016})}\BibitemShut {NoStop}%
\bibitem [{\citenamefont {Wilting}\ and\ \citenamefont
  {Priesemann}(2019{\natexlab{b}})}]{Wilting2019b}%
  \BibitemOpen
  \bibfield  {author} {\bibinfo {author} {\bibfnamefont {J.}~\bibnamefont
  {Wilting}}\ and\ \bibinfo {author} {\bibfnamefont {V.}~\bibnamefont
  {Priesemann}},\ }\href {\doibase 10.1016/j.conb.2019.08.002} {\bibfield
  {journal} {\bibinfo  {journal} {Curr. Opin. Neurobiol.}\ }\textbf {\bibinfo
  {volume} {58}},\ \bibinfo {pages} {105 } (\bibinfo {year}
  {2019}{\natexlab{b}})}\BibitemShut {NoStop}%
\bibitem [{\citenamefont {Goldstein}\ \emph {et~al.}(2004)\citenamefont
  {Goldstein}, \citenamefont {Morris},\ and\ \citenamefont
  {Yen}}]{Goldstein2004}%
  \BibitemOpen
  \bibfield  {author} {\bibinfo {author} {\bibfnamefont {M.~L.}\ \bibnamefont
  {Goldstein}}, \bibinfo {author} {\bibfnamefont {S.~A.}\ \bibnamefont
  {Morris}}, \ and\ \bibinfo {author} {\bibfnamefont {G.~G.}\ \bibnamefont
  {Yen}},\ }\href {\doibase 10.1140/epjb/e2004-00316-5} {\bibfield  {journal}
  {\bibinfo  {journal} {Eur. Phys. J. B - Condens. Matter Complex Syst.}\
  }\textbf {\bibinfo {volume} {41}},\ \bibinfo {pages} {255} (\bibinfo {year}
  {2004})}\BibitemShut {NoStop}%
\bibitem [{\citenamefont {Papanikolaou}\ \emph {et~al.}(2011)\citenamefont
  {Papanikolaou}, \citenamefont {Bohn}, \citenamefont {Sommer}, \citenamefont
  {Durin}, \citenamefont {Zapperi},\ and\ \citenamefont
  {Sethna}}]{Papanikolaou2011}%
  \BibitemOpen
  \bibfield  {author} {\bibinfo {author} {\bibfnamefont {S.}~\bibnamefont
  {Papanikolaou}}, \bibinfo {author} {\bibfnamefont {F.}~\bibnamefont {Bohn}},
  \bibinfo {author} {\bibfnamefont {R.~L.}\ \bibnamefont {Sommer}}, \bibinfo
  {author} {\bibfnamefont {G.}~\bibnamefont {Durin}}, \bibinfo {author}
  {\bibfnamefont {S.}~\bibnamefont {Zapperi}}, \ and\ \bibinfo {author}
  {\bibfnamefont {J.~P.}\ \bibnamefont {Sethna}},\ }\href {\doibase
  10.1038/nphys1884} {\bibfield  {journal} {\bibinfo  {journal} {Nat. Phys.}\
  }\textbf {\bibinfo {volume} {7}},\ \bibinfo {pages} {316} (\bibinfo {year}
  {2011})}\BibitemShut {NoStop}%
\bibitem [{\citenamefont {Friedman}\ \emph {et~al.}(2012)\citenamefont
  {Friedman}, \citenamefont {Ito}, \citenamefont {Brinkman}, \citenamefont
  {Shimono}, \citenamefont {DeVille}, \citenamefont {Dahmen}, \citenamefont
  {Beggs},\ and\ \citenamefont {Butler}}]{Beggs2012}%
  \BibitemOpen
  \bibfield  {author} {\bibinfo {author} {\bibfnamefont {N.}~\bibnamefont
  {Friedman}}, \bibinfo {author} {\bibfnamefont {S.}~\bibnamefont {Ito}},
  \bibinfo {author} {\bibfnamefont {B.~A.~W.}\ \bibnamefont {Brinkman}},
  \bibinfo {author} {\bibfnamefont {M.}~\bibnamefont {Shimono}}, \bibinfo
  {author} {\bibfnamefont {R.~E.~L.}\ \bibnamefont {DeVille}}, \bibinfo
  {author} {\bibfnamefont {K.~A.}\ \bibnamefont {Dahmen}}, \bibinfo {author}
  {\bibfnamefont {J.~M.}\ \bibnamefont {Beggs}}, \ and\ \bibinfo {author}
  {\bibfnamefont {T.~C.}\ \bibnamefont {Butler}},\ }\href {\doibase
  10.1103/PhysRevLett.108.208102} {\bibfield  {journal} {\bibinfo  {journal}
  {Phys. Rev. Lett.}\ }\textbf {\bibinfo {volume} {108}},\ \bibinfo {pages}
  {208102} (\bibinfo {year} {2012})}\BibitemShut {NoStop}%
\bibitem [{\citenamefont {Laurson}\ \emph {et~al.}(2013)\citenamefont
  {Laurson}, \citenamefont {Illa}, \citenamefont {Santucci}, \citenamefont
  {Tore~Tallakstad}, \citenamefont {M{\aa}l{\o}y},\ and\ \citenamefont
  {Alava}}]{Laurson2013}%
  \BibitemOpen
  \bibfield  {author} {\bibinfo {author} {\bibfnamefont {L.}~\bibnamefont
  {Laurson}}, \bibinfo {author} {\bibfnamefont {X.}~\bibnamefont {Illa}},
  \bibinfo {author} {\bibfnamefont {S.}~\bibnamefont {Santucci}}, \bibinfo
  {author} {\bibfnamefont {K.}~\bibnamefont {Tore~Tallakstad}}, \bibinfo
  {author} {\bibfnamefont {K.~J.}\ \bibnamefont {M{\aa}l{\o}y}}, \ and\
  \bibinfo {author} {\bibfnamefont {M.~J.}\ \bibnamefont {Alava}},\ }\href
  {\doibase 10.1038/ncomms3927} {\bibfield  {journal} {\bibinfo  {journal}
  {Nature Communications}\ }\textbf {\bibinfo {volume} {4}},\ \bibinfo {pages}
  {2927} (\bibinfo {year} {2013})}\BibitemShut {NoStop}%
\bibitem [{\citenamefont {Rybarsch}\ and\ \citenamefont
  {Bornholdt}(2014)}]{Rybarsch2014}%
  \BibitemOpen
  \bibfield  {author} {\bibinfo {author} {\bibfnamefont {M.}~\bibnamefont
  {Rybarsch}}\ and\ \bibinfo {author} {\bibfnamefont {S.}~\bibnamefont
  {Bornholdt}},\ }\href {\doibase 10.1371/journal.pone.0093090} {\bibfield
  {journal} {\bibinfo  {journal} {PLoS One}\ }\textbf {\bibinfo {volume} {9}}
  (\bibinfo {year} {2014}),\ 10.1371/journal.pone.0093090}\BibitemShut
  {NoStop}%
\bibitem [{\citenamefont {Miller}\ \emph {et~al.}(2019)\citenamefont {Miller},
  \citenamefont {Yu},\ and\ \citenamefont {Plenz}}]{MillerYuPlenz:2019}%
  \BibitemOpen
  \bibfield  {author} {\bibinfo {author} {\bibfnamefont {S.~R.}\ \bibnamefont
  {Miller}}, \bibinfo {author} {\bibfnamefont {S.}~\bibnamefont {Yu}}, \ and\
  \bibinfo {author} {\bibfnamefont {D.}~\bibnamefont {Plenz}},\ }\href
  {\doibase 10.1038/s41598-019-52326-y} {\bibfield  {journal} {\bibinfo
  {journal} {Sci. Rep.}\ }\textbf {\bibinfo {volume} {9}} (\bibinfo {year}
  {2019}),\ 10.1038/s41598-019-52326-y}\BibitemShut {NoStop}%
\bibitem [{\citenamefont {Berger}(1929)}]{Berger1929}%
  \BibitemOpen
  \bibfield  {author} {\bibinfo {author} {\bibfnamefont {H.}~\bibnamefont
  {Berger}},\ }\href {\doibase 10.1007/BF01797193} {\bibfield  {journal}
  {\bibinfo  {journal} {Arch. Psychiatr.}\ }\textbf {\bibinfo {volume} {87}},\
  \bibinfo {pages} {527} (\bibinfo {year} {1929})}\BibitemShut {NoStop}%
\bibitem [{\citenamefont {Buzs\'aki}\ and\ \citenamefont
  {Draguhn}(2004)}]{Buzsaki2004}%
  \BibitemOpen
  \bibfield  {author} {\bibinfo {author} {\bibfnamefont {G.}~\bibnamefont
  {Buzs\'aki}}\ and\ \bibinfo {author} {\bibfnamefont {A.}~\bibnamefont
  {Draguhn}},\ }\href {\doibase 10.1126/science.1099745} {\bibfield  {journal}
  {\bibinfo  {journal} {Science}\ }\textbf {\bibinfo {volume} {304}},\ \bibinfo
  {pages} {1926} (\bibinfo {year} {2004})}\BibitemShut {NoStop}%
\bibitem [{\citenamefont {Penttonen}\ and\ \citenamefont
  {Buzs\'aki}(2003)}]{Penttonen2003}%
  \BibitemOpen
  \bibfield  {author} {\bibinfo {author} {\bibfnamefont {M.}~\bibnamefont
  {Penttonen}}\ and\ \bibinfo {author} {\bibfnamefont {G.}~\bibnamefont
  {Buzs\'aki}},\ }\href {\doibase 10.1017/S1472928803000074} {\bibfield
  {journal} {\bibinfo  {journal} {Thalamus Relat. Syst.}\ }\textbf {\bibinfo
  {volume} {2}},\ \bibinfo {pages} {145} (\bibinfo {year} {2003})}\BibitemShut
  {NoStop}%
\bibitem [{\citenamefont {Kingman}(1992)}]{Kingman1992}%
  \BibitemOpen
  \bibfield  {author} {\bibinfo {author} {\bibfnamefont {J.}~\bibnamefont
  {Kingman}},\ }\href@noop {} {\emph {\bibinfo {title} {Poisson processes}}}\
  (\bibinfo  {publisher} {Clarendon Press},\ \bibinfo {address} {Oxford},\
  \bibinfo {year} {1992})\BibitemShut {NoStop}%
\bibitem [{\citenamefont {Park}(2018)}]{Park2018}%
  \BibitemOpen
  \bibfield  {author} {\bibinfo {author} {\bibfnamefont {K.~I.}\ \bibnamefont
  {Park}},\ }\href {\doibase 10.1007/978-3-319-68075-0} {\emph {\bibinfo
  {title} {Fundamentals of Probability and Stochastic Processes with
  Applications to Communications}}}\ (\bibinfo  {publisher} {Springer},\
  \bibinfo {year} {2018})\BibitemShut {NoStop}%
\bibitem [{\citenamefont {Kuntz}\ and\ \citenamefont
  {Sethna}(2000)}]{KuntzSethna:2000}%
  \BibitemOpen
  \bibfield  {author} {\bibinfo {author} {\bibfnamefont {M.~C.}\ \bibnamefont
  {Kuntz}}\ and\ \bibinfo {author} {\bibfnamefont {J.~P.}\ \bibnamefont
  {Sethna}},\ }\href {\doibase 10.1103/PhysRevB.62.11699} {\bibfield  {journal}
  {\bibinfo  {journal} {Phys. Rev. B}\ }\textbf {\bibinfo {volume} {62}},\
  \bibinfo {pages} {11699} (\bibinfo {year} {2000})}\BibitemShut {NoStop}%
\bibitem [{\citenamefont {Dobrinevski}\ \emph {et~al.}(2014)\citenamefont
  {Dobrinevski}, \citenamefont {Doussal},\ and\ \citenamefont
  {Wiese}}]{DobrinevskiLeDoussalWiese:2014}%
  \BibitemOpen
  \bibfield  {author} {\bibinfo {author} {\bibfnamefont {A.}~\bibnamefont
  {Dobrinevski}}, \bibinfo {author} {\bibfnamefont {P.~L.}\ \bibnamefont
  {Doussal}}, \ and\ \bibinfo {author} {\bibfnamefont {K.~J.}\ \bibnamefont
  {Wiese}},\ }\href@noop {} {\bibfield  {journal} {\bibinfo  {journal}
  {Europhys. Lett.}\ }\textbf {\bibinfo {volume} {108}},\ \bibinfo {pages}
  {66002} (\bibinfo {year} {2014})}\BibitemShut {NoStop}%
\bibitem [{\citenamefont {Baldassarri}\ \emph {et~al.}(2003)\citenamefont
  {Baldassarri}, \citenamefont {Colaiori},\ and\ \citenamefont
  {Castellano}}]{BaldassarriColaioriCastellano:2003}%
  \BibitemOpen
  \bibfield  {author} {\bibinfo {author} {\bibfnamefont {A.}~\bibnamefont
  {Baldassarri}}, \bibinfo {author} {\bibfnamefont {F.}~\bibnamefont
  {Colaiori}}, \ and\ \bibinfo {author} {\bibfnamefont {C.}~\bibnamefont
  {Castellano}},\ }\href@noop {} {\bibfield  {journal} {\bibinfo  {journal}
  {Phys. Rev. Lett.}\ }\textbf {\bibinfo {volume} {90}},\ \bibinfo {pages}
  {060601} (\bibinfo {year} {2003})}\BibitemShut {NoStop}%
\bibitem [{\citenamefont {Willis}\ and\ \citenamefont
  {Pruessner}(2018)}]{WillisPruessner:2018}%
  \BibitemOpen
  \bibfield  {author} {\bibinfo {author} {\bibfnamefont {G.}~\bibnamefont
  {Willis}}\ and\ \bibinfo {author} {\bibfnamefont {G.}~\bibnamefont
  {Pruessner}},\ }\href@noop {} {\bibfield  {journal} {\bibinfo  {journal}
  {Int. J. Mod. Phys.}\ }\textbf {\bibinfo {volume} {32}},\ \bibinfo {pages}
  {1830002} (\bibinfo {year} {2018})}\BibitemShut {NoStop}%
\bibitem [{\citenamefont {Doi}(1976)}]{Doi:1976}%
  \BibitemOpen
  \bibfield  {author} {\bibinfo {author} {\bibfnamefont {M.}~\bibnamefont
  {Doi}},\ }\href@noop {} {\bibfield  {journal} {\bibinfo  {journal} {J. Phys.
  A: Math. Gen.}\ }\textbf {\bibinfo {volume} {9}},\ \bibinfo {pages} {1465}
  (\bibinfo {year} {1976})}\BibitemShut {NoStop}%
\bibitem [{\citenamefont {Peliti}(1985)}]{Peliti:1985}%
  \BibitemOpen
  \bibfield  {author} {\bibinfo {author} {\bibfnamefont {L.}~\bibnamefont
  {Peliti}},\ }\href@noop {} {\bibfield  {journal} {\bibinfo  {journal} {J.
  Phys. (Paris)}\ }\textbf {\bibinfo {volume} {46}},\ \bibinfo {pages} {1469}
  (\bibinfo {year} {1985})}\BibitemShut {NoStop}%
\bibitem [{\citenamefont {Pausch}(2019)}]{Pausch2019}%
  \BibitemOpen
  \bibfield  {author} {\bibinfo {author} {\bibfnamefont {J.}~\bibnamefont
  {Pausch}},\ }\emph {\bibinfo {title} {Topics in Statistical Mechanics}},\
  \href@noop {} {Ph.D. thesis},\ \bibinfo  {school} {Imperial College London}
  (\bibinfo {year} {2019}),\ \bibinfo {note}
  {http://hdl.handle.net/10044/1/73905}\BibitemShut {NoStop}%
\end{thebibliography}%
